\documentclass[preprint,prd,nofootinbib,preprintnumbers]{revtex4}
\usepackage{inputenc}
\usepackage{epsfig,float}
\usepackage{amsfonts}
\usepackage{amsmath}
\usepackage{longtable}
\usepackage{bm}
\usepackage[colorlinks=true,linkcolor=red,citecolor=blue]{hyperref}

\input{epsf}
\newcommand{\be}{\begin{eqnarray}}
\newcommand{\ee}{\end{eqnarray}}
\newcommand{\ba}{\begin{array}}
\newcommand{\ea}{\end{array}}

\newcommand{\bi}{\begin{itemize}}
\newcommand{\ei}{\end{itemize}}
\newcommand{\nn}{{\nonumber}}

\restylefloat{figure}
\restylefloat{table}

\begin{document}

\title{Backward timelike Compton scattering to decipher the photon content of the nucleon}

\author{Bernard~Pire$^1$,  Kirill~M.~Semenov-Tian-Shansky$^{2,3}$,
Alisa~A. Shaikhutdinova$^{2,3}$,  Lech~Szymanowski$^{4}$ }
\affiliation{
$^1$ CPHT, CNRS, \'{E}cole Polytechnique, I.P. Paris,  91128 Palaiseau, France  \\
$^2$ National Research Centre Kurchatov Institute: Petersburg Nuclear Physics Institute, 188300 Gatchina, Russia \\
$^3$ Higher School of Economics,
National Research University, 194100 St. Petersburg, Russia \\
$^4$ National Centre for Nuclear Research, NCBJ, 02-093 Warsaw, Poland
}

\preprint{CPHT-RR003.012022}

\begin{abstract}
The exclusive photoproduction off nucleon of a large invariant mass lepton pair in the backward region specified by the small Mandelstam variable $-u$
is discussed in the framework of collinear QCD factorization. The amplitude is factorized in terms of photon-to-nucleon Transition Distribution Amplitudes (TDAs) which encode the  photon content of the nucleon.  A simplistic model-estimate of   these new non-perturbative objects is used to estimate the magnitude of the corresponding cross sections. The background due to the electromagnetic Bethe-Heitler process is shown to be negligible in the kinematical regime of interest.

\end{abstract}

\maketitle
\thispagestyle{empty}
\renewcommand{\thesection}{\arabic{section}}
\renewcommand{\thesubsection}{\arabic{subsection}}

\section{Introduction}

Generalized Parton Distributions (GPDs)
\cite{Muller:1994ses}
are since more than 20 years a subject of intense theoretical study and experimental effort.
The process which is most profoundly studied among those where GPDs factorize from
perturbatively calculable coefficient functions, is Deeply Virtual Compton Scattering (DVCS)
$\gamma^*(q) N(p_N) \rightarrow \gamma(q') N(p_N')$,
see {\it e.g.} reviews
\cite{Diehl:2003ny,Belitsky:2005qn}.
It involves a spacelike virtual photon
$\gamma^*(q)$
with
$q^2<0$,
providing a hard scale of this process with a small absolute value of
$t$-channel invariant momentum transfer
$t=(p'-p)^2$.
The ``inverse'' process,
$\gamma(q) N(p_N) \rightarrow \gamma^*(q') N(p_N')$
with large timelike virtuality
$q'^{2}>0$ of the final state photon
$\gamma^*(q')$,
{\it i.e.} timelike Compton scattering (TCS)
\cite{Berger:2001xd},
shares many common features with DVCS. The hard scale in this case is the final state photon virtuality
$q'^{2}$, and one still considers the kinematical region defined by small absolute value of
$t=(p_N'-p_N)^2$.
The studies of TCS permit to test the universality of GPDs entering both DVCS and TCS,
which is one of the main consequences of the QCD collinear factorization theorem.
Moreover, the crossing from a spacelike to a timelike probe  provides an important test of the understanding of QCD corrections \cite{Pire:2011st, Moutarde:2013qs} and of the analytic structure of the amplitude \cite{Mueller:2012sma}.
The first experimental study of TCS performed  at JLab was recently reported by the CLAS collaboration
\cite{CLAS:2021lky}.

The near-backward TCS (bTCS) is a TCS process occurring in a different kinematical region, which is characterized by
a small absolute value of the Mandelstam variable $u=(p_N'-q)^2$ and
a large negative value of the variable  $t=(p_N'-p_N)^2$.
Thus bTCS involves exchanges in the $u-$channel carrying baryonic quantum number and its factorized scattering amplitude contains a new non-perturbative object, the photon-to-nucleon transition distribution amplitude (TDA), see Fig. \ref{Fig_TDAfact}. TDAs extend the concepts of GPDs and baryon DAs, see the review \cite{Pire:2021hbl}. The ultimate goal of the study of TDAs is to uncover new facets of the understanding of the nucleon wave function. In the present case,
photon-to-nucleon TDAs provide a key to decipher the photon content of the nucleon.

The case of backward DVCS was preliminary discussed in \cite{Pire:2004ie, Lansberg:2006uh}. It involves the much related nucleon-to-photon TDAs which we will discuss together with the photon-to-nucleon TDAs. We specialize here to the TCS process which seems to us much easier to analyze experimentally because of the large contamination of $\pi^0 N$ final states in a backward DVCS study.
After a detailed presentation of kinematics in Sec.~\ref{Sec_Kinematics}, we discuss the properties of photon-to-nucleon TDAs in Sec.~\ref{Sec_gamma_to_N_TDA}. We calculate the bTCS scattering amplitude and cross-section in Sec.~\ref{Sec_Scat_Ampl_Bkw_TCS}.  In Sec.~\ref{Sec_BH}, we calculate the Bethe-Heitler contribution to the bTCS process and show why we can safely ignore it in an experimental study. In Sec.~\ref{estimate}, we give a simplistic estimate of near backward TCS cross-section using an oversimplified model of photon to nucleon TDAs based on vector dominance. In Sec.~\ref{Sec_CS_eN} we estimate the expected  rates for a quasi-real photon experiment at an electron beam facility. Sec.~\ref{Sec_Concl} briefly presents our conclusions.

\section{Kinematics of near-backward timelike Compton scattering }
\label{Sec_Kinematics}

We consider the timelike Compton scattering
\be
\gamma(q)+N(p_N,s) \to
\gamma^*(q', \lambda_\gamma)+N(p'_N,s')   \to
\ell^+(k_{\ell^+}) \ell^-(k_{\ell^-})+N(p'_N,s').
\label{TCS_reaction}
\ee
Throughout this paper we employ the usual Mandelstam variables for the
hard subprocess of the reaction
(\ref{TCS_reaction}):
$$
s=(p_N+q)^2 \equiv W^2; \ \ \ t=(p'_N-p_N)^2; \ \ \ u=(q'-p_N)^2 \equiv \Delta^2.
$$

The
$\gamma N $
center-of-mass energy squared
$s=(p_N+q)^2 \equiv W^2$
and the virtuality of the final state timelike photon
$q'^2=Q'^2$
introduce the natural hard scale.
We define the analogue of the Bjorken variable
\be
\tau \equiv \frac{Q'^2}{2 p_N \cdot q}= \frac{Q'^2}{W^2-m_N^2}.
\label{Def_tau}
\ee

\begin{figure}[h]
 \centering
\includegraphics[width=0.35\textwidth]{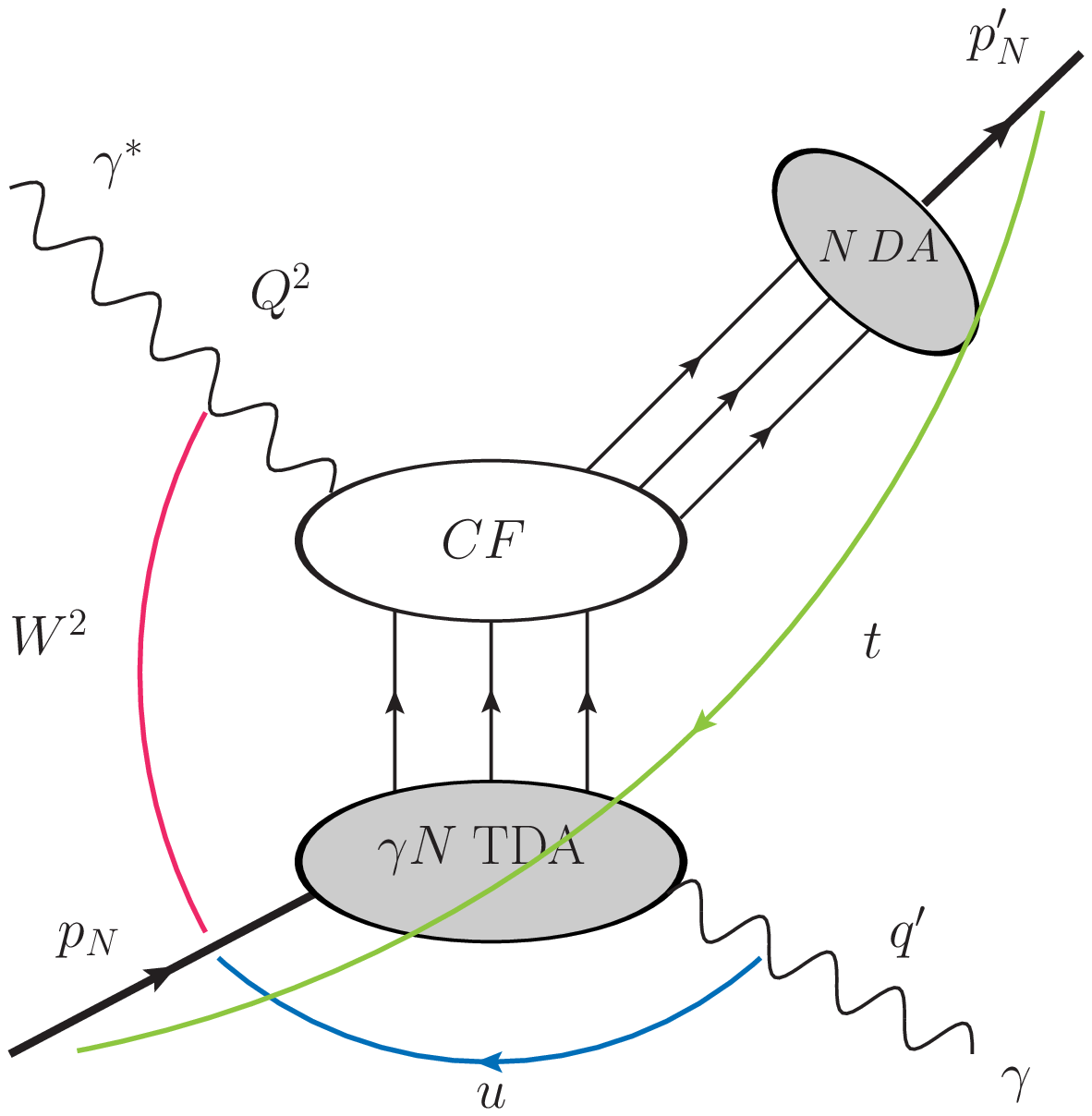}  \ \ \ \ \ \ \ \ \ \ \ \ \
\includegraphics[width=0.45\textwidth]{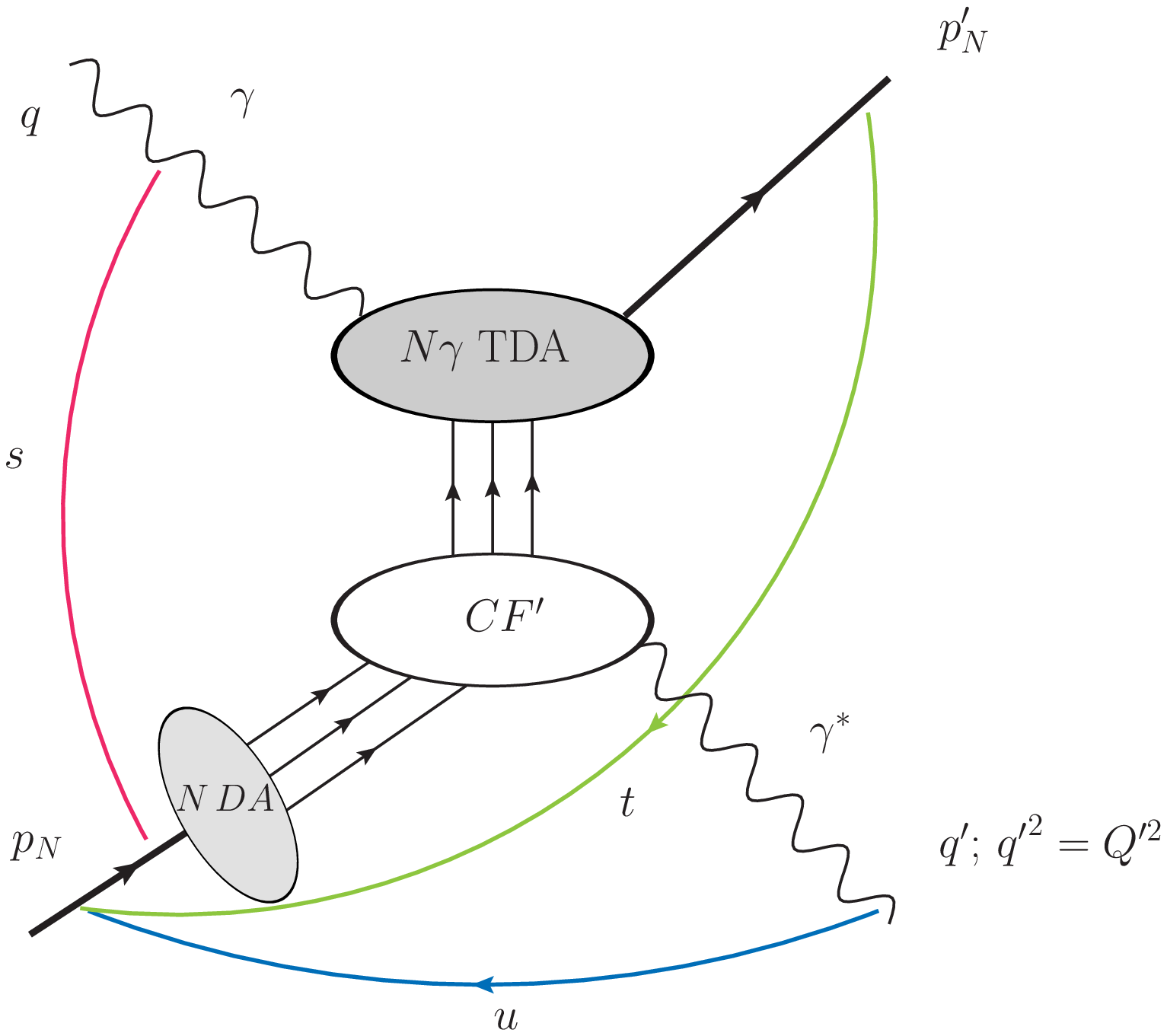}
    \caption{ {\bf Left panel:} Collinear factorization mechanism  for DVCS
      ($\gamma^* N \to N' \gamma$) in the  near-backward  kinematical regime (large $-q^2 \equiv Q^2$, $W^2 \equiv (p_N+q)^2$; fixed $x_{B}=\frac{Q^2}{2 p_N \cdot q}$;
      $|u| \equiv |(p_N-q')^2| \sim 0$).
    {\bf Right panel:}
    Collinear factorization  of TCS ($\gamma N \to \gamma^* N' $)  in the  near-backward  kinematical regime (large $q'^2 \equiv Q'^2$, $W^2 \equiv (p_N+q)^2 $; fixed $\tau$ (\ref{Def_tau});   $|u|\equiv |(q'-p_N)^2| \sim 0$);
  $\gamma N$ ($N\gamma$) TDA stands for the transition distribution amplitudes from a nucleon-to-a-photon (photon-to-a-nucleon); $N$ DA stands for the nucleon distribution amplitude;
      $CF$ and $CF'$ denote hard subprocess amplitudes (coefficient functions).  }
\label{Fig_TDAfact}
\end{figure}

In complete analogy with our
analysis of the nucleon-antinucleon annihilation process
\cite{Lansberg:2007se,Lansberg:2012ha},
we assume that this reaction admits a factorized description  in the
near-backward kinematical regime,
where
$Q'^2$ and $W^2$ are large; $\tau$ is fixed; and
the $u$-channel momentum transfer squared is small compared to $Q'^2$ and $W^2$:
$|u| \equiv |\Delta^2|= |(p'_N-q)^2| \ll W^2, \, Q'^2$.
Within such kinematics, the amplitude of the hard subprocess of the reaction
(\ref{TCS_reaction})
is supposed to admit a collinear factorized description in terms of
photon-to-nucleon TDAs and nucleon DAs, as  shown on the right panel of Fig.~\ref{Fig_TDAfact}.
Small $|u|$ corresponds to the kinematics where the virtual photon is produced in the near-backward direction, in the
$\gamma(q) N(p_N)$ center-of-mass system (CMS).
Therefore, in what follows we refer the kinematical regime in question as the near-backward kinematics.
This kinematical regime is complementary to the more familiar
regime of TCS
($Q'^2$ and $W^2$ - large; $\tau$ -fixed; $|t| \ll Q'^2,\,s$),
known as the near-forward kinematics.
In this latter kinematical regime the conventional collinear factorization theorem
leading to the description of the reaction
(\ref{TCS_reaction})
in terms of GPDs is established
\cite{Muller:1994ses,Berger:2001xd}
and is favored by recent experimental results
\cite{CLAS:2021lky}.

We choose the $z$-axis along the colliding real-photon-nucleon
and introduce the light-cone vectors
$p$
and
$n$
($p^2=n^2=0$; $2 p \cdot n=1$).

We consider  the following Sudakov decomposition for the momenta of the reaction
(\ref{TCS_reaction})
in the near-backward kinematical regime:
\be
&&
q=(1+\xi) p; \nn \\ &&
p_{N}= \frac{m_N^2 (1+\xi)}{W^2- m_N^2} p  
+
\frac{W^2-m_N^2}{1+\xi }
 n; \nn \\  &&
\Delta \equiv \left(p'_{N}-q\right)=-2 \xi p+\left(\frac{m_{N}^{2}-\Delta_{T}^{2}}{1-\xi}
\right) n+\Delta_{T};
\nn \\ &&
q'=p_N-\Delta ; \quad p'_{N}=q+\Delta.
\label{Sudakov_dec_omega}
\ee
Here
$m_N$
is the nucleon mass and the skewness variable $\xi$ is defined with
respect to the longitudinal momentum transfer between the initial
state real photon and the final state nucleon
\be
\xi \equiv  -\frac{(p'_N-q) \cdot n}{(p'_N+q) \cdot n}.
\label{Def_xi}
\ee
Within the collinear factorization framework we neglect both the squared
nucleon mass with respect to
$q'^2=Q'^2$
and
$W^2$
and set
$\Delta_T=0$
within the leading twist coefficient function.
This results in the approximate expression for the  skewness variable
(\ref{Def_xi}):
\be
\xi \simeq \frac{Q'^2}{2 W^2-Q'^2} \simeq \frac{\tau}{2 -\tau}.
\label{Xi_collinear}
\ee

The approximation
(\ref{Xi_collinear})
can potentially affect the definition of the physical
domain of the reaction
(\ref{TCS_reaction})
determined by the requirement
$\Delta_T^2 \le 0$,
where
\be
\Delta_T^2= \frac{1-\xi}{1+\xi} \left( u+
\frac{2\xi m_N^2}{1-\xi} \right).
\label{Def_DT2}
\ee

It is also instructive to consider the exact kinematics of the reaction
(\ref{TCS_reaction})
in the
$\gamma N$
CMS frame. In this frame the relevant momenta read:
\be
&&
q= \left(     \frac{W^2-m_N^2}{2W}    , \, \vec{q}\right);
\ \ \ \ \
q'= \left(  \frac{W^2+Q'^2-m_N^2}{2W}, \, -\vec{p}\,'_N \right); \nonumber
\\ &&
p_N= \left( \frac{W^2+m_N^2}{2W} , \, -\vec{q} \right);
\ \ \ \ \
p'_N= \left(  \frac{W^2+m_N^2-Q'^2}{2W}, \, \vec{p}\,'_N \right),
\ee
where
\be
|\vec{q}|= \frac{W^2-m_N^2}{2W}; \ \ \
|\vec{p}\,'_N|=\frac{\Lambda(W^2,m_N^2,Q'^2)}{2W}\,,
\ee
with
\be
\Lambda(x,y,z)= \sqrt{x^2+y^2+z^2-2xy-2xz-2yz}
\label{Def_Mandelstam_f}
\ee
being the Mandelstam function.

The CMS scattering angle
$\theta_u^*$
is defined as the angle between
$\vec{q}$
and
$\vec{p}\,'_N$:
\be
\cos \theta_u^*= \frac{2W^2(u-m_N^2)+(W^2-m_N^2)(W^2+m_N^2-Q'^2)}
{(W^2-m_N^2)\Lambda(W^2,m_N^2,Q'^2)}.
\label{CosTheta_exact}
\ee
The transverse momentum transfer squared
(\ref{Def_DT2})
is then given by
\be
\Delta_T^2=- \frac{\Lambda^2(W^2,Q'^2,m_N^2)}{4W^2}(1-\cos^2 \theta_u^*)\,,
\label{DeltaT2_exact}
\ee
and the physical domain for the reaction
(\ref{TCS_reaction})
is defined from the requirement that
$\Delta_T^2 \le 0 $.

\bi
\item
In particular, the backward kinematics regime
$\theta_u^*=0$
corresponds to
$\vec{p}\,'_N$
along
$\vec{q}$,
which means that the virtual photon
$\gamma^*$
is produced along
$-\vec{q}$ {\it i.e.}
in the near-backward direction. In this case
$u$
reaches its maximal value
\be
u_{0} \equiv \frac{2 \xi m_N^2(\xi+1)}{\xi^2-1}
\nonumber 
=m_N^2- \frac{(W^2-m_N^2)(W^2+m_N^2-Q'^2)}{2W^2}
+2|\vec{q}||\vec{p}\,'_N|.
\label{Def_umax}
\ee
At the same time,
$t=(p'_N-p_N)^2$
reaches its minimal value
$t_{1}$
\be
t_{1}=2m_N^2+Q'^2-W^2-u_{0}.
\ee

It is for
$u \sim u_{0}$
that one may expect to satisfy the requirement
$|u| \ll W^2, \, Q'^2$
that is crucial for the validity of the factorized description of
(\ref{TCS_reaction})
in terms of
$\gamma \to N$
TDAs and nucleon DAs.

\item The other limiting value
 $\theta_u^*=\pi$
corresponds to
$\vec{p}\,'_N$ along $-\vec{q}$
{\it i.e}
$\gamma^*$ is
produced in the near-forward direction.
In this case $u$ reaches its minimal value
\be
u_{1}=m_N^2- \frac{(W^2-m_N^2)(W^2+m_N^2-Q'^2)}{2W^2}
-2|\vec{q}||\vec{p}\,'_N|.
\ee
At the same time
$t$
reaches its maximal value
$t_{0}$.
The factorized description in terms of
$\gamma \to N$
TDAs does not apply in this case as
$|u|$
turns out to be of the order of
$W^2$. This is the usual domain of application of the GPD factorization.
\ei

\section{The photon-to-nucleon TDAs}
\label{Sec_gamma_to_N_TDA}

The collinear factorized description of the near-backward TCS
involves photon-to-nucleon ($N \gamma$) TDAs defined as the
Fourier transform of the $N \gamma$ matrix element of the trilocal
light-cone operator%
\footnote{We assume the use of the light-cone gauge $A^+ \equiv 2 (A \cdot n)=0$ and omit the Wilson lines along the light-like path.}
\begin{equation}
\hspace{2em}\widehat{O}_{\rho \tau \chi}^{\bar u \bar u \bar d}(\lambda_1n,\lambda_2n,\lambda_3n) =\varepsilon_{c_1 c_2 c_3}
 \bar u^{c_1}_{\rho}(\lambda_1 n)
\bar u^{c_2}_{\tau}(\lambda_2 n)\bar d^{c_3}_{\chi}(\lambda_3 n).
\end{equation}
Here $c_{1,2,3}$ stand for the color group indices; and $\rho$, $\tau$, $\chi$
denote the Dirac indices.

A crossing relation links photon-to-nucleon TDAs to the already extensively discussed nucleon-to-photon TDAs
\cite{Pire:2004ie,Lansberg:2006uh}
(see also Sec.~4.1.3 of Ref.~\cite{Pire:2021hbl}), which enter the
collinear factorized description of the near-backward DVCS (see left panel of Fig.~\ref{Fig_TDAfact}).
This crossing relation  is similar to that between nucleon-to-pion
($\pi N$)
and pion-to-nucleon
($N \pi$)
TDAs
established in Ref.~\cite{Pire:2016gut}
in the context of  the QCD-based description of near-backward production of a
$J/\psi$ off nucleons with a pion beam.

To the leading twist-$3$ accuracy, the parametrization of photon-to-nucleon TDAs
involves $16$ independent $N \gamma$ TDAs
$V_{\Upsilon}^{N \gamma}, \, A_{\Upsilon}^{N \gamma}, \, T_{\Upsilon}^{N \gamma}$:
\begin{eqnarray}
&&
4 {\cal F} \langle  N^p(p_N,s_N)| \widehat{O}_{\rho \tau \chi}^{\bar u \bar u \bar d}(\lambda_1n,\lambda_2n,\lambda_3n)|\gamma(q, \lambda_\gamma) \rangle
\nonumber \\ &&
\hspace{2em}
= \delta(x_1+x_2+x_3-2\xi)  \times m_N \times \\
&&\Big[
\sum_{\Upsilon= 1 {\cal E}, 1T, \atop 2 {\cal E}, 2T  } (v^{N \gamma }_\Upsilon)_{\rho \tau, \, \chi} V_{\Upsilon}^{N \gamma}(x_1,x_2,x_3, \xi, \Delta^2; \, \mu^2)
\nonumber \\ &&
+\sum_{\Upsilon= 1 {\cal E}, 1T,  \atop 2 {\cal E}, 2T  } (a^{N \gamma}_\Upsilon)_{\rho \tau, \, \chi} A_{\Upsilon}^{N \gamma}(x_1,x_2,x_3, \xi, \Delta^2; \, \mu^2)\nonumber\\&&
+
\sum_{\Upsilon= 1 {\cal E}, 1T,    2{\cal E}, 2T, \atop 3 {\cal E}, 3T,  4 {\cal E}, 4T } (t^{N \gamma}_\Upsilon)_{\rho \tau, \, \chi} T_{\Upsilon}^{N \gamma}(x_1,x_2,x_3, \xi, \Delta^2; \, \mu^2)
\Big]\,,
\nonumber
\label{Def_N_gamma_TDAs_param}
\end{eqnarray}
with the Fourier transform operation
$4 {\cal F} \equiv 4(p \cdot n)^{3} \int\left[\prod_{j=1}^{3} \frac{d \lambda_{j}}{2 \pi}\right] e^{-i \sum_{k=1}^{3} x_{k} \lambda_{k}(p \cdot n)}$.
Each of these $16$
$N \gamma$
TDAs is a function of the light-cone momentum fractions $x_{1,2,3}$;
the skewness variable
$\xi$
(\ref{Def_xi}), the $u$-channel invariant momentum transfer
$\Delta^2$;
and the factorization scale
$\mu^2$.

The explicit form of the crossing relation between the $N \gamma$ and $\gamma N$ TDAs
reads (see Appendix~A of Ref.~\cite{Pire:2016gut}):
\be
\left\{V_{\Upsilon}^{N \gamma}, \, A_{\Upsilon}^{N \gamma}, \, T_{\Upsilon}^{N \gamma}\right\}\left(x_{1,2,3}, \xi, \Delta^{2}\right)
=
\left\{V_{\Upsilon}^{\gamma N }, \, A_{\Upsilon}^{\gamma N }, \, T_{\Upsilon}^{\gamma N }\right\}
\left(-x_{1,2,3},-\xi, \Delta^{2}\right).
\label{Crossing_prescription}
\ee
The set of the leading twist-$3$ Dirac structures
$s_{\rho \tau, \chi}^{
N \gamma
}= \{v^{N \gamma }_\Upsilon, \, a^{N \gamma }_\Upsilon, \,
t^{N \gamma }_\Upsilon \}_{\rho \tau, \chi}$
occurring in
(\ref{Def_N_gamma_TDAs_param})
is related to
$s_{\rho \tau, \chi}^{
  \gamma N
}= \{v^{\gamma N }_\Upsilon, \, a^{\gamma N }_\Upsilon, \,
t^{\gamma N }_\Upsilon \}_{\rho \tau, \chi}$
entering the parametrization of $\gamma N$ TDAs listed in Appendix~\ref{App_B} as
\be
s_{\rho \tau, \chi}^{
N \gamma}
=
\left(\gamma_{0}^{T}\right)_{\tau \tau^{\prime}}\left[s_{\rho^{\prime} \tau^{\prime}, \chi^{\prime}}^{
\gamma N
}\right]^{\dagger}\left(\gamma_{0}\right)_{\rho^{\prime} \rho}\left(\gamma_{0}\right)_{\chi^{\prime} \chi}.
\ee

The counting of the independent
leading twist $ N \gamma$ TDAs matches the number of independent helicity amplitudes
$T_{h_{\bar u} h_{\bar u},h_{\bar d}}^{ h_\gamma , h_N }$
for $\gamma \,  u u d \to N^p$ process, where
$h_{\bar q}$, $h_\gamma$ and $h_N$
refer to the light-cone helicity of, respectively, quark, initial state photon and the
final state nucleon.

In the strictly backward limit $\Delta_T=0$ only $4$ $N \gamma$
TDAs $V_{1 {\cal E}}^{N \gamma}$, $A_{1 {\cal E}}^{N \gamma}$, $T_{1 {\cal E}}^{N \gamma}$,
$T_{2 {\cal E}}^{N \gamma}$ are relevant.
This is consistent with the number of helicity conserving processes (with obvious helicity notations:  $\uparrow : h=+1 ; \downarrow : h=-1 $):
\be
&&
\gamma(\uparrow) u(\uparrow)u(\downarrow)d(\downarrow) \to N(\uparrow); \nn \\ &&
\gamma(\uparrow) u(\downarrow)u(\uparrow)d(\downarrow) \to N(\uparrow); \nn \\ &&
\gamma(\uparrow) u(\downarrow)u(\downarrow)d(\uparrow) \to N(\uparrow); \nn \\ &&
\gamma(\downarrow) u(\uparrow)u(\uparrow)d(\uparrow) \to N(\uparrow);
\ee
since
$\Delta_T=0$ implies helicity conservation.

In this paper we mainly concentrate on the very small $ \Delta_T$ region
where the contribution of remaining 12 $N  \gamma$ TDAs is small.
The relations between the four TDAs relevant for
$\Delta_T=0$
and light-front helicity amplitudes read:
\begin{eqnarray}
\hspace{1em}V_{1 {\cal E}}^{N \gamma }&= \frac{1}{2^{1 / 4} \sqrt{1+\xi}\left(P^{+}\right)^{3 / 2}} \frac{1}{m_N}\left[T_{\uparrow \downarrow,\uparrow}^{\uparrow  \uparrow}+T_{\downarrow \uparrow,\uparrow}
^{\uparrow   \uparrow}\right]; \nonumber\\
A_{1 {\cal E}}^{ N \gamma }&=- \frac{1}{2^{1 / 4} \sqrt{1+\xi}\left(P^{+}\right)^{3 / 2}}
 \frac{1}{m_N}\left[T_{\uparrow \downarrow,\uparrow}^{\uparrow  \uparrow}-T_{\downarrow \uparrow,\uparrow}^{\uparrow \uparrow}\right];\\
T_{1 {\cal E}}^{N \gamma }&=-\frac{1}{2^{1 / 4} \sqrt{1+\xi}\left(P^{+}\right)^{3 / 2}}
\frac{1}{m_{N}}\left[T_{\downarrow \downarrow,\uparrow}^{\uparrow \uparrow}+T_{\uparrow \uparrow,\uparrow}^{\downarrow \uparrow}\right];\nonumber \\
T_{2 {\cal E}}^{ N \gamma }&=-\frac{1}{2^{1 / 4} \sqrt{1+\xi}\left(P^{+}\right)^{3 / 2}} \frac{1}{m_{N}}\left[T_{\downarrow \downarrow,\uparrow}^{\uparrow \uparrow}-T_{\uparrow \uparrow,\uparrow}^{\downarrow \uparrow}\right],\nonumber
\end{eqnarray}
where $P\equiv \frac{q+p'_N}{2}$ is the mean momentum.
Similar relations exist for the other TDAs, which do not contribute much to the cross-sections for
$\Delta_T \simeq 0$
(see the discussion in Sec.~\ref{estimate}
around Fig. \ref{Fig_CSVMD}).

\begin{figure}[h]
 \begin{center}
\includegraphics[width=.48\textwidth]{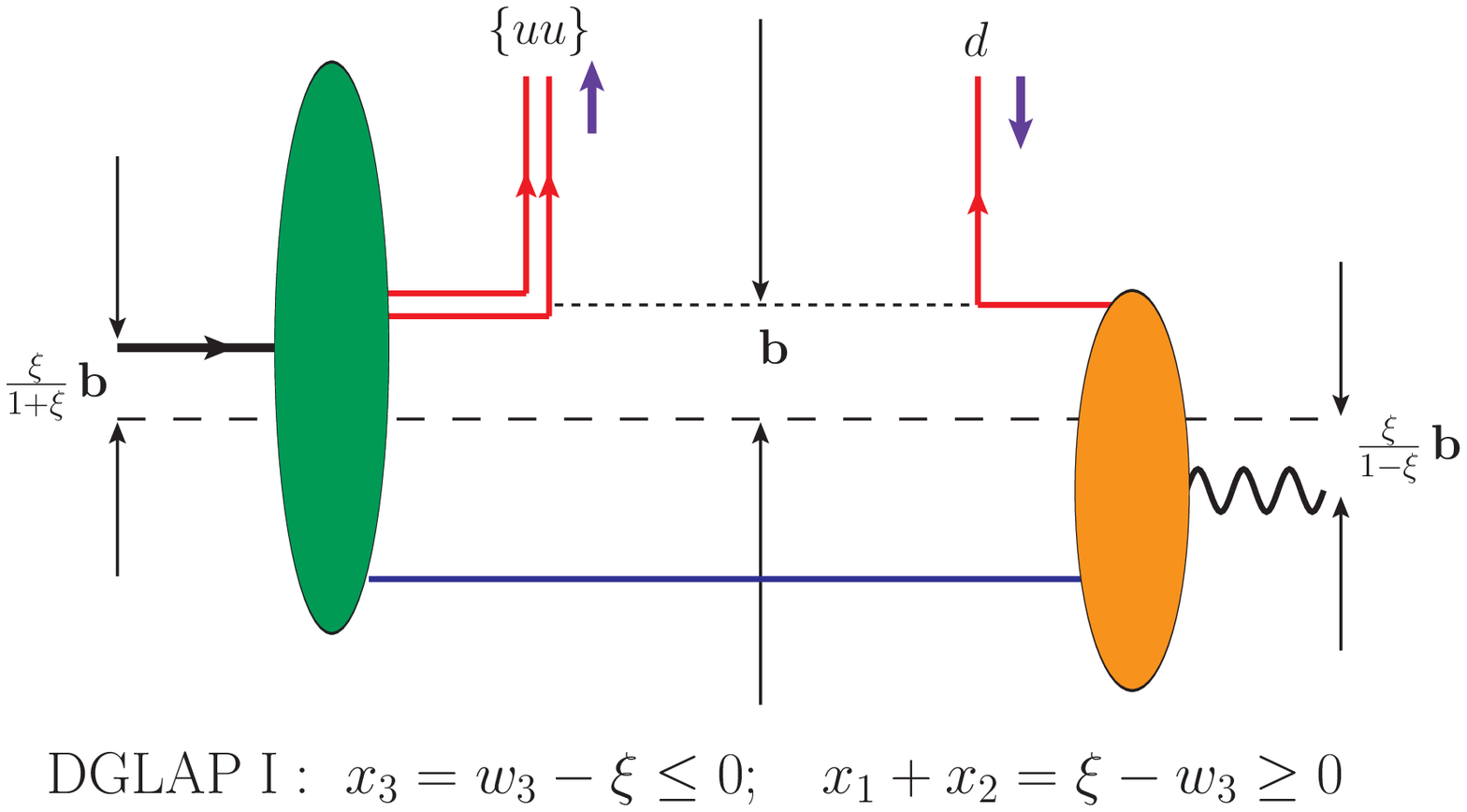}
\includegraphics[width=.48\textwidth]{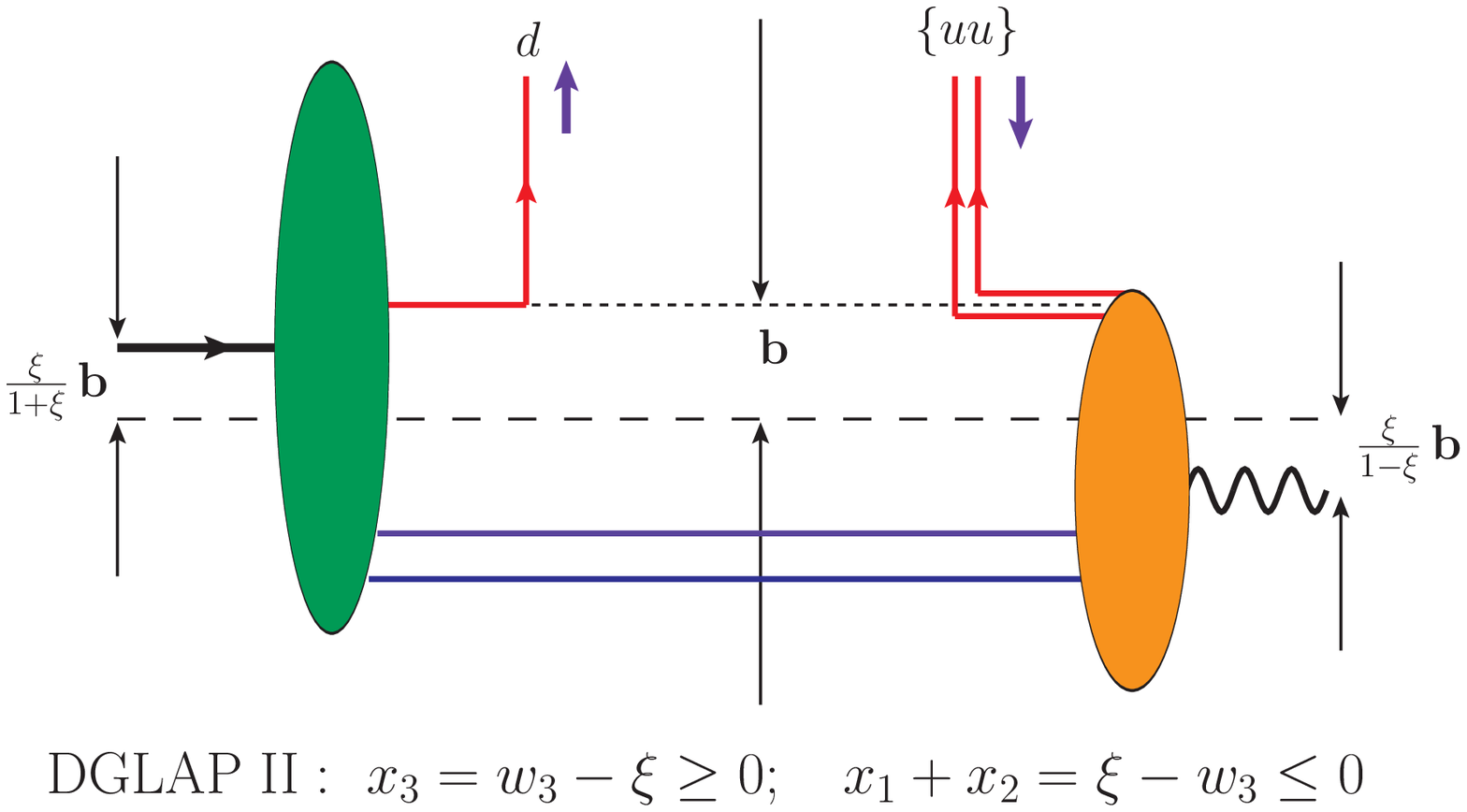} \\
  \vspace{1em}
\includegraphics[width=.48\textwidth]{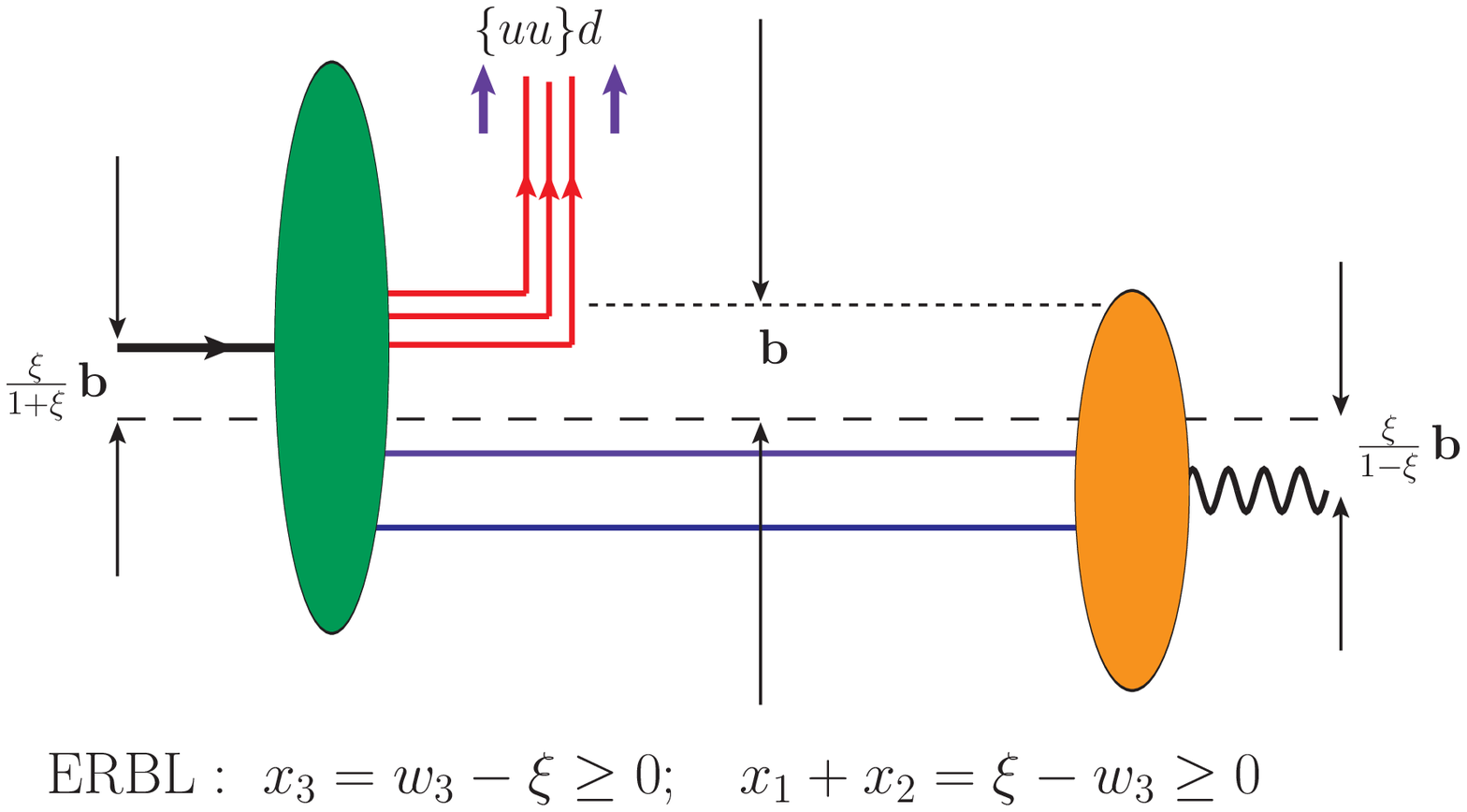}
\end{center}
\caption{Impact parameter space interpretation for the $v_3$-integrated $uud$ proton-to-photon TDA in the DGLAP-like~I, DGLAP-like~II  and in the ERBL-like domains. Solid arrows show the direction of the positive longitudinal momentum flow. }
\label{Fig_Impact}
\end{figure}

As for the nucleon-to-meson TDAs case
\cite{Pire:2019nwa}, the physical contents of
nucleon-to-photon TDAs can be addressed within
an impact parameter picture.
For that issue it is convenient to switch to the
set of so-called quark-diquark coordinates, {\it e.g.} $w_3 \equiv x_3-\xi$;
$v_3 \equiv \frac{x_1-x_2}{2}$ and consider the
Fourier transform to the impact parameter
$\Delta_T \to b_T$
of $v_3$-integrated nucleon-to-photon TDAs.
This leads to a quark-diquark interpretation in the impact parameter
space
(see Fig.~\ref{Fig_Impact})
that is largely analogous to the  impact parameter interpretation of GPDs.

\bi
\item In the DGLAP-like~I region $w_3 \le -\xi$ the impact parameter
specifies the location where a $uu$-diquark is pulled out of a proton and
then replaced by an antiquark $\bar{d}$ to annihilate into the final state photon.
\item In the DGLAP-like~II region $w_3 \ge \xi$ the impact parameter
specifies the location where a quark $d$ is pulled out of a proton and
then replaced by an antidiquark $\bar{u}\bar{u}$ which forms into the final state photon.
\item In the ERBL-like region $-\xi \le w_3 \le \xi$ the impact parameter
specifies the location where a three-quark cluster composed of a $uu$-diquark
and a $d$-quark is pulled out of the initial nucleon in a way that the remnants form the final state photon.
\ei

By considering a different choice of quark-diquark coordinates one can obtain complementary
impact parameter interpretation for $v_1 \equiv \frac{x_2-x_3}{2}$-integrated and
$v_2 \equiv \frac{x_3-x_1}{2}$-integrated nucleon-to-photon TDAs with different pairs of
quarks chosen to form a diquark inside nucleon. Therefore, the study of the impact parameter
representation of $v$-integrated nucleon-to-photon TDAs allows to address the question:
{\em Where in the transverse plane does the nucleon emit a photon}?

\section{The scattering amplitude and the backward TCS cross-section}
\label{Sec_Scat_Ampl_Bkw_TCS}

Let us now calculate the scattering amplitude of the process
\be
&&
\gamma(q, \lambda_\gamma) + N(p_N,s_N) \to \gamma^*(q', \lambda'_\gamma)+ N'(p'_N,s'_N)
\nn \\&&
\to  N'(p'_N,s'_N)+\ell^+(k_{\ell^+},s_{\ell^+}) +\ell^-(k_{\ell^-},s_{\ell^-})\,,
\label{React}
\ee
which contains the TDAs just defined in Section~\ref{Sec_gamma_to_N_TDA}.

The  helicity amplitudes  ${\cal M}^{\lambda_\gamma \lambda'_\gamma}_{s_N s'_N}$ of the hard subprocess reaction
\be
\gamma(q, \lambda_\gamma) + N(p_N,s_N) \to \gamma^*(q', \lambda'_\gamma)+ N'(p'_N,s'_N)
\label{HardReact}
\ee
involve $4$ independent tensor structures
\be
{\cal M}_{s_N s'_N}^{\lambda_\gamma \lambda'_\gamma}=
{\cal C}
\frac{1}{Q'^4}
\sum_{k=1,3,4,5}
{\cal S}_{s_N s'_N}^{(k) \, \lambda_\gamma \lambda_V} {\cal J}_{\rm \, bTCS}^{(k)} (\xi, \, \Delta^2).
\label{Hel_ampl_def}
\ee

Here
${\cal J}_{\rm \, bTCS}^{(k)}$, $k=1,3,4,5$ stand for the integral convolutions of $N\gamma$ TDAs and nucleon DAs
with hard scattering kernels.
The normalization  constant
\be
{\cal C}=
-i\left(\frac{\left(4 \pi \alpha_{s}\right)^{2} \sqrt{4 \pi \alpha_{\mathrm{em}}} f_{N} m_N}{54}\right),
\label{Def_C}
\ee
where $\alpha_{\mathrm{em}} \simeq \frac{1}{137}$ ($\alpha_s \simeq 0.3$) is the electromagnetic (strong) coupling; and
$f_N$
is the nucleon light-cone wave function normalization constant.
We will use
$
f_N=5.0 \cdot 10^{-3} \; {\rm GeV}^2
$~\cite{Chernyak:1987nv}.

There is one tensor structure independent of $\Delta_T$:
\be
&&
{\cal S}_{s_N s'_N}^{(1) \, \lambda_\gamma \lambda'_\gamma}=
\bar{U}(p'_N,s'_N) \hat{\cal E}^*(q',\lambda'_\gamma) \hat{\cal E}(q, \lambda_\gamma) U(p_N,s_N)\,,
\ee
and three
$\Delta_T$-dependent tensor structures:
\be
&&
{\cal S}_{s_N s'_N}^{(3) \, \lambda_\gamma \lambda'_\gamma}= \frac{1}{m_N} ({\cal E}(q, \lambda_\gamma) \cdot \Delta_T) \, \bar{U}(p'_N,s'_N) \hat{{\cal E}}^*(q',\lambda'_\gamma)  U(p_N,s_N);
\nonumber \\ &&
{\cal S}_{s_N s'_N}^{(4) \, \lambda_\gamma \lambda'_\gamma}=\frac{1}{m_N^2} ({\cal E}(q, \lambda_\gamma) \cdot \Delta_T) \, \bar{U}(p'_N,s'_N) \hat{{\cal E}}^*(q',\lambda') \hat{\Delta}_T U(p_N,s_N);
\nonumber \\
&&
{\cal S}_{s_N s'_N}^{(5)\,\lambda_\gamma \lambda'_\gamma}= \bar{U}(p'_N,s'_N) \hat{{\cal E}}^*(q',\lambda'_\gamma) \hat{\cal E}^*(q, \lambda_\gamma) \hat{\Delta}_T U(p_N,s_N)\,.
\ee

To the leading order in $\alpha_s$, the hard amplitude is calculated from the $21$ Feynman diagrams analogous to those familiar from the calculation of nucleon electromagnetic form factor (see {\it e.g.} Ref.~\cite{Chernyak:1984bm}).
To the leading twist accuracy, the integral convolutions of  $N\gamma$ TDAs and nucleon DAs
${\cal J}^{(k)}_{\rm \, bTCS}$, $k=1,\,3,\,4,\,5$ with hard scattering kernels $T_{\alpha}^{(k)}$
read
\be
&&
\mathcal{J}_{\rm \, bTCS}^{(k)}  (\xi,\Delta^2) =
{\int^{1+\xi}_{-1+\xi} }\! \! \!d_3x  \; \delta \left( \sum_{j=1}^3 x_j-2\xi \right)
{\int^{1}_{0} } \! \! \! d_3y \; \delta \left( \sum_{l=1}^3 y_l-1 \right) \;
{\Bigg(2\sum_{\alpha=1}^{7}   T_{\alpha}^{(k)} +
\sum\limits_{\alpha=8}^{14}   T_{\alpha}^{(k)} \Bigg)}, \nn \\ &&
\label{Def_I_k_convolutions}
\ee
where the index $\alpha$ refers to the number of the diagram.
The explicit expressions for $T_{\alpha}^{(k)}$
are listed in Table~I of Appendix~\ref{App_A}.

Therefore, to the leading twist accuracy the averaged-squared amplitude
(\ref{Average_sq_Ampl})
integrated
over the lepton azimuthal angle
$\varphi_\ell$
reads
\be
\int d \varphi_\ell \overline{|{\cal M}_{N \gamma \to N \ell^+ \ell^-}|^2} \Big|_{\text{Leading twist-3}}=
|\overline{{\cal M}_T}|^2 \frac{2 \pi e^2 (1+  \cos^2 \theta_\ell)}{Q'^2},
\label{M2_leading_tw}
\ee
where
\be
&&
|\overline{{\cal M}_T}|^2 =
\frac{1}{4}
\sum_{s_N s'_N {\lambda  \lambda'}}
{\cal M}^{\lambda \lambda'}_{s_N s'_N}
\left( {\cal M}^{\lambda \lambda'}_{s_N s'_N} \right)^* 
=\frac{1}{4} |{\cal C}|^2 \frac{1}{Q'^6}
\frac{2(1+\xi}{\xi}
\Big[
2|{\cal J}^{(1)}_{\rm \, bTCS}|^2 \nn \\ && + \frac{\Delta_T^2}{m_N^2} \Big\{
-|{\cal J}^{(3)}_{\rm \, bTCS}|^2+ \frac{\Delta_T^2}{m_N^2} |{\cal J}^{(4)}_{\rm \, bTCS}|^2+
\left( {\cal J}^{(4)}_{\rm \, bTCS} {\cal J}^{(1)*}_{\rm \, bTCS}+{\cal J}^{(4)*}_{\rm \, bTCS} {\cal J}^{(1)}_{\rm \, bTCS}\right) \nn \\ &&
-2 |{\cal J}^{(5)}_{\rm \, bTCS}|^2 -
\left( {\cal J}^{(5)}_{\rm \, bTCS} {\cal J}^{(3)*}_{\rm \, bTCS}+{\cal J}^{(5)*}_{\rm \, bTCS} {\cal J}^{(3)}_{\rm \, bTCS}\right)
\Big\} + {\cal O}(1/Q'^2)
\Big].
\label{TransAmpl_squared_Cross}
\ee

The
$2 \to 3$ particle scattering
differential cross section of (\ref{React}) reads (see {\it e.g.} \cite{Borodulin:2017pwh}):
\be
&&
d \sigma= \frac{(2 \pi)^4}{2 \Lambda(s,m_N^2,0)}
\overline{|{\cal M}_{N \gamma \to N' \ell^+ \ell^-}|^2} \nn \\ &&
\times
\delta^{(4)}(p_N+p_V-p'_N-k_{\ell^+}-k_{\ell^-})
\frac{d^3 p'_N}{(2 \pi)^3 2 E_{N'}}
\frac{d^3 k_{\ell^+}}{(2 \pi)^3 2 E_{\ell^+}}
\frac{d^3 k_{\ell^-}}{(2 \pi)^3 2 E_{\ell^-}},
\ee
where $\Lambda$
is the  Mandelstam function (\ref{Def_Mandelstam_f}).

\begin{figure}[H]
\begin{center}
\includegraphics[width=0.75\textwidth]{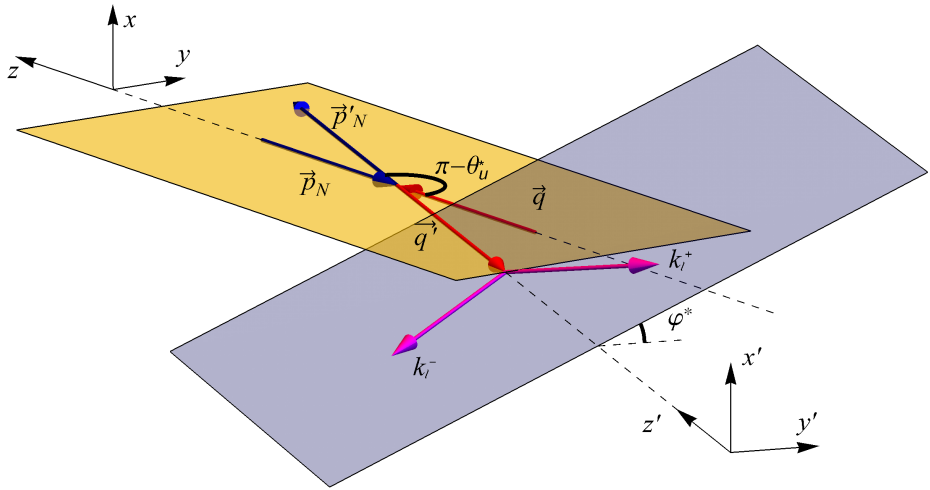}
\end{center}
     \caption{The kinematical variables and coordinate axes in the $\gamma p$  CMS.
     The CMS scattering angle $\theta_u^*$ (\ref{CosTheta_exact})
is defined as the angle between
$\vec{q}$
and
$\vec{p}\,'_N$.
     The boost
     along the $z'$ direction allows to switch to $\ell^+ \ell^-$-CMS frame. The azimuthal angle $\varphi^*$ is defined as the angle between the hadronic and leptonic planes.  The lepton polar angle $\theta_\ell$ is
     defined between $\vec{k}_{\ell^+}$ and $\vec{p}\,'_N$ in the $\ell^+ \ell^-$-CMS frame. The lepton azimuthal angle $\varphi_\ell$ can be chosen to coincide with $\varphi^*$ since a boost along $z'$ does not change the momentum components transverse with respect to $z'$.}
\label{Fig_KinX}
\end{figure}

The $3$-particle Lorentz-invariant phase space (LIPS) is decomposed in the usual manner
into two  $2$-particle LIPSs of $\gamma^* N'$ and $\ell^+ \ell^-$ systems.
We then get (neglecting lepton masses):
\be
d \sigma = \frac{1}{2 (2 \pi)^5 \Lambda(s,m_N^2,0)}
\overline{|{\cal M}_{N \gamma \to N' \ell^+ \ell^-}|^2}
\frac{d \Omega_{N'}^*}{8 s} \Lambda(s,m_N^2,Q'^2) \frac{d \Omega_\ell}{8} dQ'^2,
 \ee
where $d \Omega_{N'} \equiv d \cos \theta_{u}^* d \varphi^*$
is the produced nucleon solid angle in the $\gamma N$ CMS (see Fig.~\ref{Fig_KinX}); and
$d \Omega_{\ell}$ is the solid angle of the produced lepton in the $\ell^+ \ell^-$
CMS.
We express $\cos \theta_{u}^*$ through $u=(q-p'_N)^2$:
\be
du= \frac{d \cos \theta_{u}^*}{2 s}
(s-m_N^2)
\Lambda(s,m_N^2,Q'^2).
\ee
Then, finally,
\be
\frac{d \sigma}{du d Q'^2 d \cos \theta_{\ell}}=
\frac{\int d \varphi_\ell \overline{|{\cal M}_{N \gamma \to N' \ell^+ \ell^-}|^2} }{64
(s-m_N^2)^2
(2 \pi)^4}\,,
\label{CS_main_formula}
\ee
where the average-squared amplitude
$\overline{|{\cal M}_{N \gamma \to N' \ell^+ \ell^-}|^2}$
is expressed as
\be
\overline{|{\cal M}_{N \gamma \to N' \ell^+ \ell^-}|^2}=
\frac{1}{4}
\sum_{s_N s'_N {\lambda_\gamma  \lambda'_\gamma}} \frac{1}{Q'^4} e^2
{\cal M}^{\lambda_\gamma \lambda'_\gamma}_{s_N s'_N}
{\rm Tr}
\left\{
\hat{k}_{\ell^-} {\cal E}(q', \lambda'_\gamma) \hat{k}_{\ell^+} {\cal E}^*(q', \lambda'_\gamma)
\right\}
\left( {\cal M}^{\lambda_\gamma \lambda'_\gamma}_{s_N s'_N} \right)^*,
\label{Average_sq_Ampl}
\ee
where $ {\cal E}^*(q', \lambda'_\gamma)$
stands for the polarization vector of the outgoing virtual photon;
and the factor $\frac{1}{4}$ corresponds to averaging over the polarizations
of the initial nucleon and the two transverse polarizations of the initial photon.

\section{The Bethe-Heitler contribution}
\label{Sec_BH}

\begin{figure}[H]
 \begin{center}
 \includegraphics[width=0.3\textwidth]{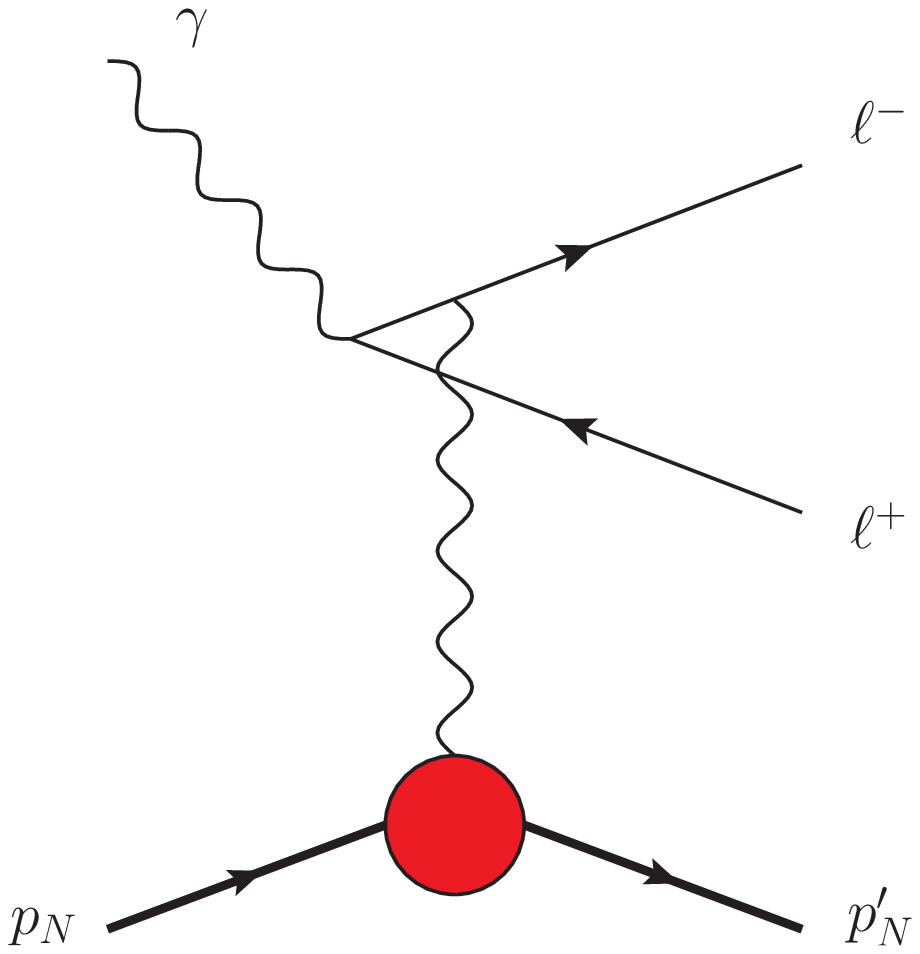} \ \ \ \ \
  \includegraphics[width=0.3\textwidth]{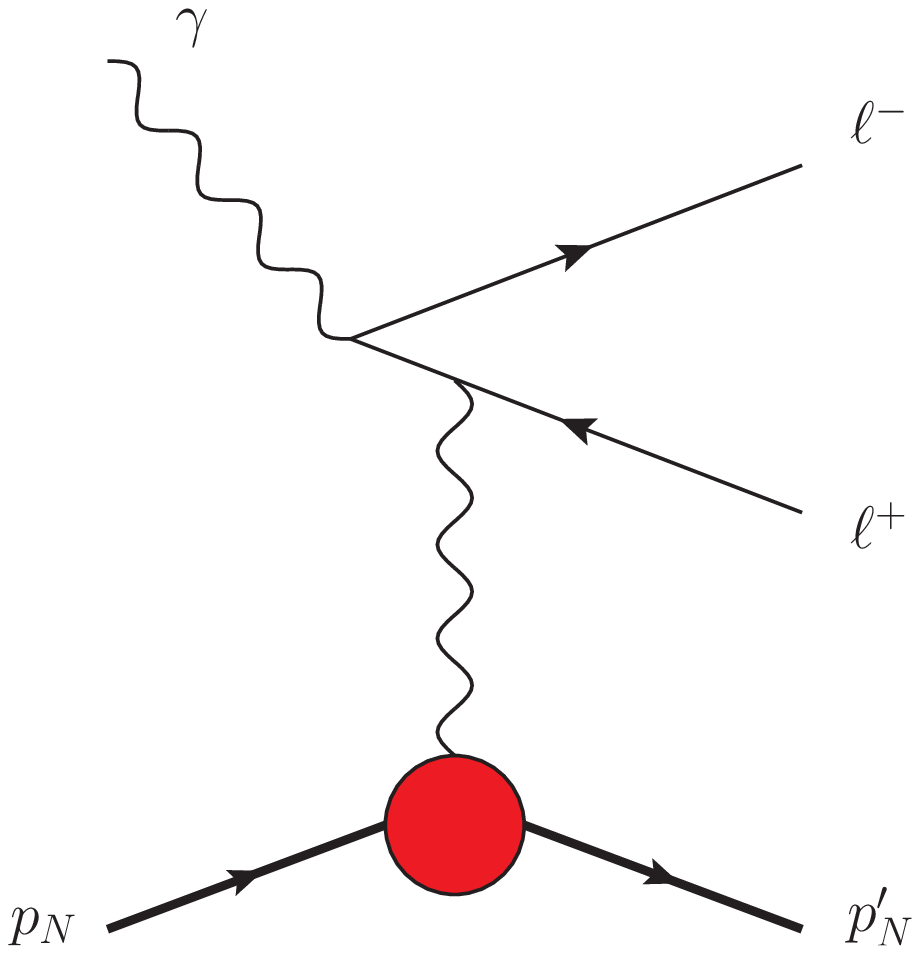}
 \end{center}
  \caption{ The  Bethe Heitler process for exclusive photoproduction of a lepton pair on a nucleon. The red disk represents the electromagnetic nucleon form factors $F_{1,2}(t)$; $-t$ is small in the near-forward kinematics but large in the near-backward kinematics studied here.}
\label{BH}
\end{figure}

As in the forward case, the process (\ref{TCS_reaction}) receives a contribution from the purely electromagnetic reaction, the Bethe-Heitler process (BH)  where the initial photon couples directly to one of the produced leptons; see Fig. \ref{BH}.

The cross-section
$\frac{d \sigma_{B H}}{d Q^{\prime 2} d t d\cos \theta_\ell d \varphi_\ell}$
for this subprocess is described in full details in Eq.~(17) of \cite{Berger:2001xd}. A careful examination of this formula demonstrates how much the forward and backward cases are different. While the intrinsic order of magnitude of the  BH cross section is very small, it may be enhanced in two kinematical regions where some inverse propagators become tiny. These regions are:
\begin{itemize}
    \item the near-forward region where one photon propagator proportional to $\frac{1}{t}$ enhances the cross section when it increases up to  $\frac{1}{t_{0}}$;
    \item one particular region of the phase space in the intermediate region where the product of the lepton inverse propagators
        $$L = \frac{(Q'^2-t)^2}{4}- \left((k_{\ell^+}-k_{\ell^-})\cdot(p_N-p'_N)\right)^2$$
        becomes small.
\end{itemize}
\begin{figure}[H]
 \begin{center}
 \includegraphics[width=0.47\textwidth]{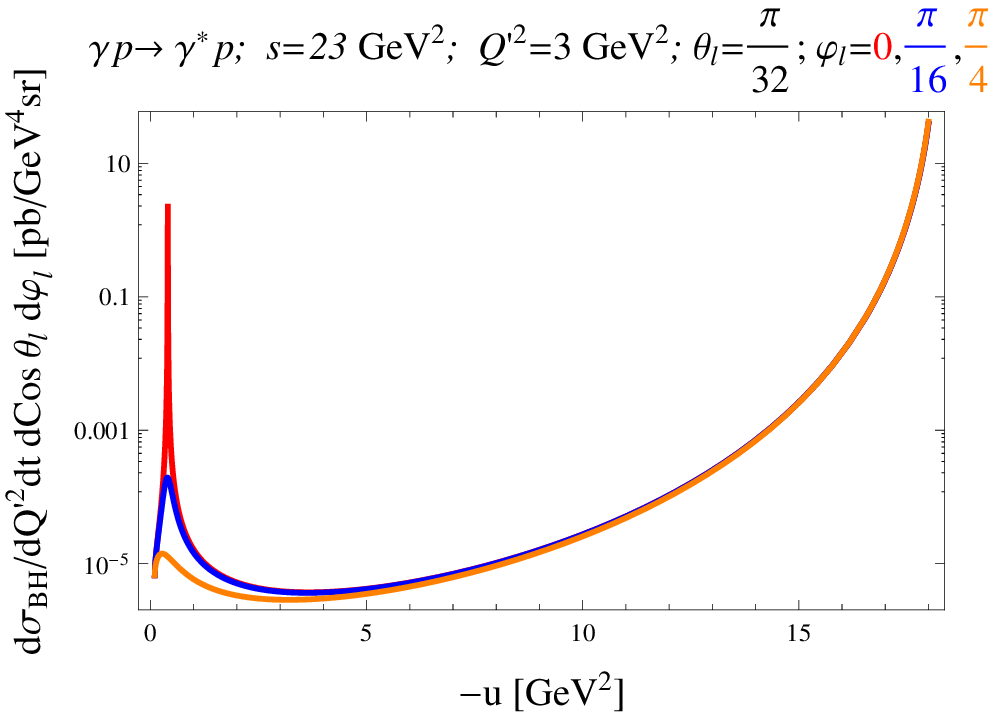} ~~~~~ \includegraphics[width=0.47\textwidth]{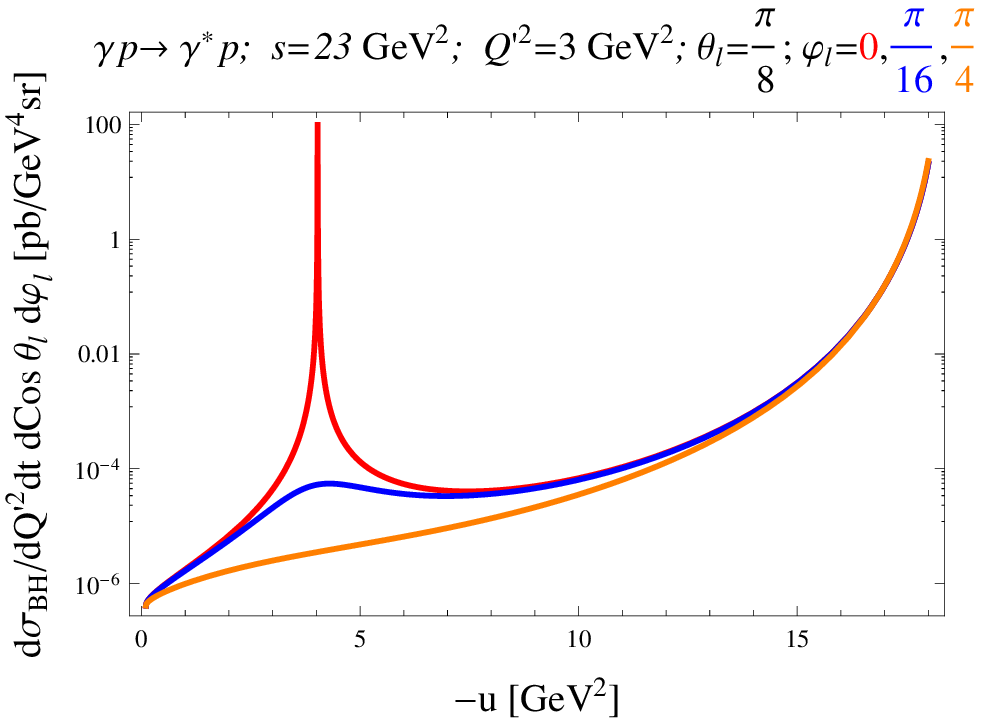}
 \end{center}
  \caption{ The $-u$ dependence of the Bethe-Heitler differential cross-section at $\theta_\ell = \pi/32$ (left panel) and $\theta_\ell = \pi/8$ (right panel) for various values of $\varphi_\ell$, at $Q'^2 = 3$ GeV$^2$ and $s=23$ GeV$^2$.}
\label{BHphi}
\end{figure}

\begin{figure}[H]
 \begin{center}
 \includegraphics[width=0.47\textwidth]{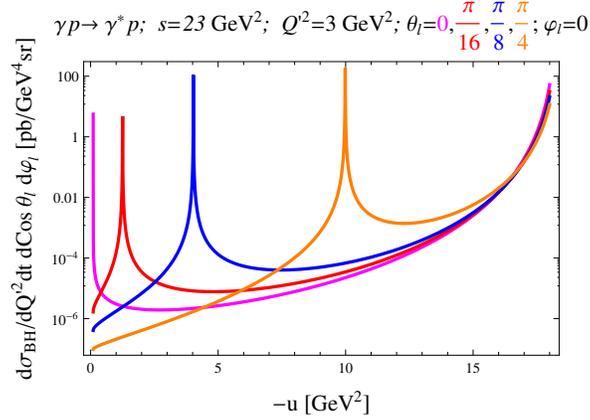}
 \end{center}
  \caption{  The $-u$ dependence of the Bethe-Heitler differential cross-section at $\varphi_\ell =0$ and different values of $\theta_\ell$ (from left to the right: $\theta_\ell=0,\,\frac{\pi}{16},\,\frac{\pi}{8},\,\frac{\pi}{4}$), at $Q'^2 = 3$~GeV$^2$ and $s=23$~GeV$^2$.}
\label{BHcs}
\end{figure}
This is illustrated in Fig. \ref{BHphi} and in Fig.~\ref{BHcs} where we plot the  Bethe-Heitler differential cross section $\frac{d \sigma_{B H}}{d Q^{\prime 2} d t d\cos \theta_\ell d \varphi_\ell}$ as a function of $-u$.
We employ the dipole representation for the nucleon form factors $F_{1,2}(t)$
and present the cross section for various values of the polar and azimuthal angles of the lepton
pair $\theta_\ell$ and $\varphi_\ell$ angles.
Apart from the peaks, the order of magnitude of the cross section is indeed tiny in the near-backward region. Analyzing the near-backward TCS cross section outside of the dangerous regions does not look like a difficult challenge.

In conclusion, easy to implement cuts on the final leptons kinematics will allow us to study the backward TCS reaction without taking into account of  the BH contribution, neither of its interference with the QCD amplitude. The necessary  cut-off concerns very small lepton emission angles when one stays far from the fixed angle region where $-u$ and $-t$ are of the same order. This reinforces our confidence that the pQCD-based reaction mechanism of
Fig.~\ref{Fig_TDAfact}
for bTCS  considerably dominates over the BH contribution in the near-backward kinematics.

\section{A simplistic estimate of near backward TCS cross-section}
\label{estimate}
Theoretical estimates of the near-backward TCS cross section
based on the QCD collinear factorization reaction mechanism require
phenomenologically complete models for the photon-to-proton TDAs.
These models do not exist yet and we definitely need experimental data to guide model builders. Since any experimental proposal needs an order of magnitude estimate of counting rates, hence of differential cross sections, we propose now an oversimplified model for photon to nucleon TDAs based on the quite usual vector dominance hypothesis often applied to photoproduction reactions. Needless to say, the physics of the TDAs is much richer than such a model, as shown in the previous sections, for instance in its ability to perform a tomography of the electromagnetic content of the nucleon. This VMD model should thus be taken as a strawman which we expect will trigger dedicated experiments which will help to extract in a second step realistic TDAs.
\subsection{A VMD model for photon to nucleon TDAs}
\label{Sec_VMD_model_calc}

In this section we present a phenomenological model for $\gamma N$
TDAs based on the Vector Meson Dominance (VMD) assumption and
the phenomenologically successful cross channel nucleon exchange model
for nucleon-to-vector meson TDAs. Relying on the
crossing relation (\ref{Crossing_prescription}) between $\gamma N$  and $N \gamma$ TDAs we then will be able to provide
the theoretical estimates for  the near-backward TCS cross section
that can be further employed for feasibility studies at the kinematical
conditions of existing and future experimental facilities.

Let us consider within the factorization framework involving
nucleon-to-vector meson TDAs (\cite{Pire:2015kxa})
the transverse amplitude of the
hard production of transversely polarized vector meson off nucleon%
\footnote{For definiteness we consider the case of the proton target.}
\be
\gamma^*(q,\lambda_\gamma)+N^p(p_N,s_N) \to V_T(p_V, \lambda_V)+N^{p'}(p'_N,s').
\label{V_T_reaction}
\ee
Here $V_T$ is a transversely polarized vector meson ($\rho^0$, $\omega$ or $\phi$). The VMD
assumption
\cite{Hakioglu:1991pn,Schildknecht:2005xr}
allows then to connect the near-backward DVCS amplitude
${\cal M}(\gamma^*N^p \to N^{p'} \gamma)$
to those of hard near-backward production of transversely polarized vector meson reactions
(\ref{V_T_reaction}):
\be
&&
{\cal M}(\gamma^*N^p \to N^{p'} \gamma) \nn  \\ && =
\frac{e}{f_{\rho}} {\cal M}(\gamma^*N^p \to N^{p'} \rho^0_T)
+\frac{e}{f_{\omega}} {\cal M}(\gamma^*N^p \to N^{p'} \omega_T)
+\frac{e}{f_{\phi}} {\cal M}(\gamma^*N^p \to N^{p'} \phi_T), 
\ee
where the vector meson-to-photon couplings
$f_{\rho,\omega,\phi}$
can be estimated from
$\Gamma_{V \to e^+e^-}$ decay widths according to \cite{Dumbrajs:1983jd}:
$\Gamma_{V \to e^+e^-}= \frac{1}{3} \alpha_{\rm em}^2 m_V \frac{4 \pi}{f_V}$.

We define the helicity amplitudes of the reaction
(\ref{V_T_reaction})
${\cal M}^{\lambda_\gamma \lambda_V}_{s_N s'_N}$
and consider
the square of the transverse amplitude $\gamma^*_T N^p \to V_T^0 N^{p'}$:
\be
&&
|\overline{{\cal M}_T}|^2 =
\frac{1}{2}
\sum_{s_N s'_N { \lambda_{\gamma_T} \lambda_{V_T}}}
{\cal M}^{\lambda_\gamma \lambda_V }_{s_N s'_N}
\left( {\cal M}^{\lambda_\gamma \lambda_V }_{s_N s'_N} \right)^* \nn \\ &&
=\frac{1}{2} |{\cal C}|^2 \frac{1}{Q^6}
\frac{2(1+\xi)}{\xi}
\Big[
2|{\cal I}^{(1)}|^2+ \frac{\Delta_T^2}{m_N^2} \Big\{
-|{\cal I}^{(3)}|^2+ \frac{\Delta_T^2}{m_N^2} |{\cal I}^{(4)}|^2+
\left( {\cal I}^{(4)} {\cal I}^{(1)*}+{\cal I}^{(4)*} {\cal I}^{(1)}\right) \nn \\ &&
-2 |{\cal I}^{(5)}|^2 -
\left( {\cal I}^{(3)} {\cal I}^{(3)*}+{\cal I}^{(5)*} {\cal I}^{(3)}\right)
\Big\} + {\cal O}(1/Q^2)
\Big].
\label{TransAmpl_squared}
\ee
Here ${\cal C}$ is defined in
(\ref{Def_C})
and ${\cal I}^{(k)}$, $k=1,3,4,5$
stand for the integral convolutions of $VN$ TDAs and nucleon DAs with
hard scattering kernels (see eq. (6.27) and Table~3 of Ref.~\cite{Pire:2021hbl}).

The summation in (\ref{TransAmpl_squared}) stands over the transverse polarizations
of the virtual photon and over the transverse polarizations of the final state vector
meson. The eq.~(\ref{TransAmpl_squared}) is to be compared with eq.~(7.21) of Ref.~\cite{Pire:2021hbl},
that gives the transverse amplitude squared summed over both transverse and longitudinal polarizations
of the final state vector meson.
Note that $16$ out of $24$ leading-twist-$3$ $VN$ TDAs contribute to the $\gamma^*_T N \to V_T N'$ amplitude.
This is consistent with the helicity counting for  $\gamma N$ TDAs.

\begin{figure}[H]
\begin{center}
\includegraphics[width=0.4\textwidth]{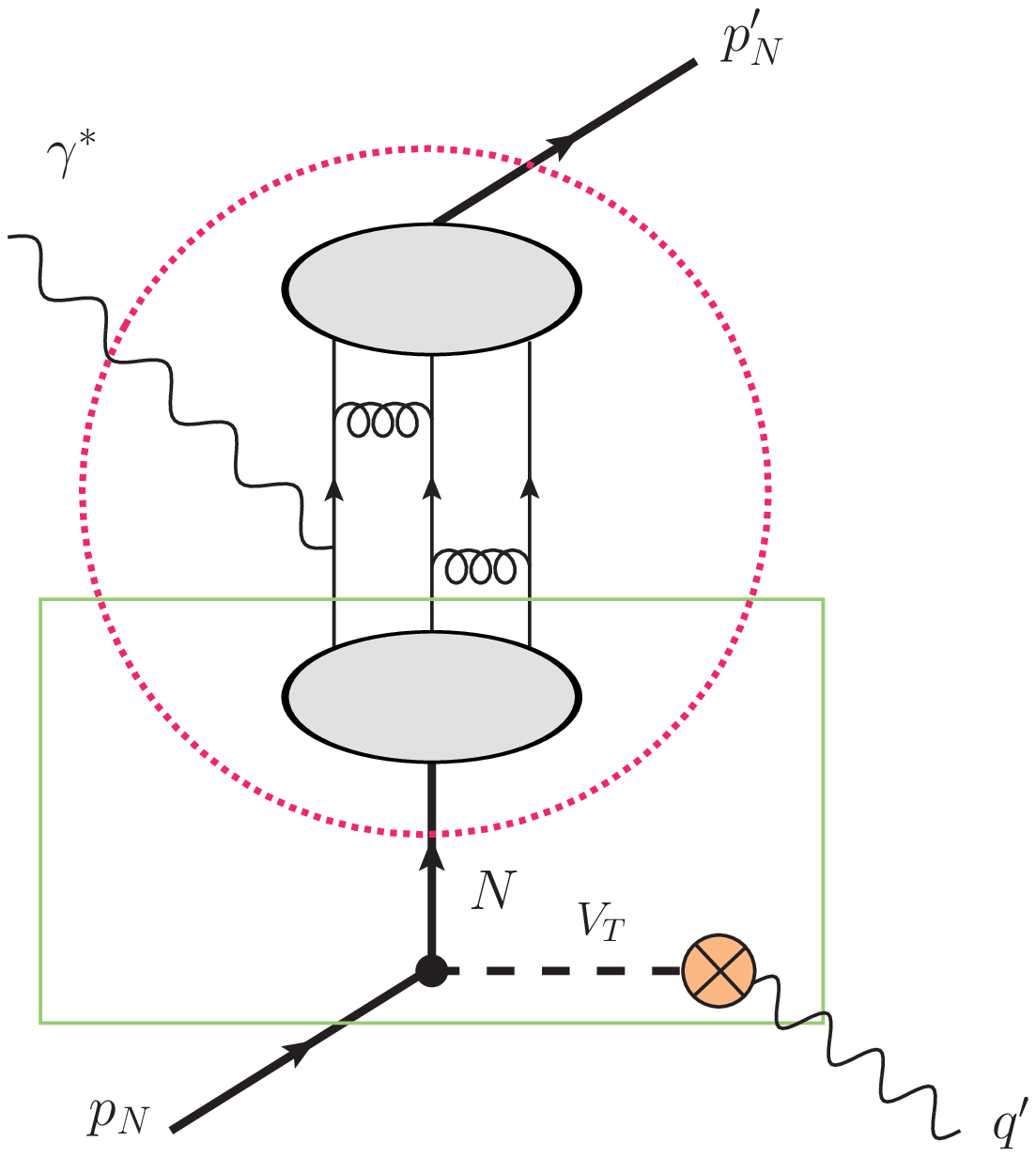}
\includegraphics[width=0.4\textwidth]{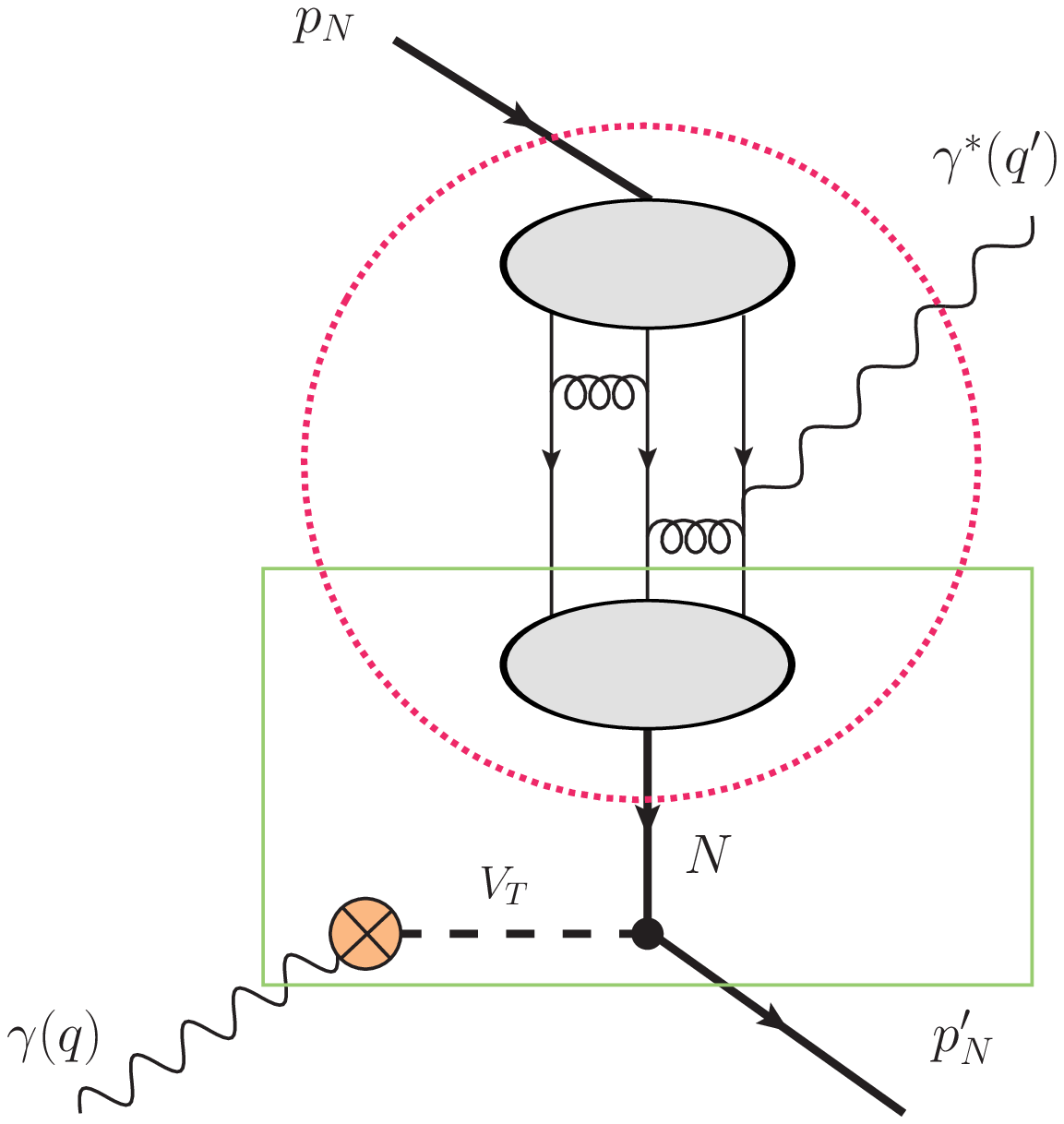}
\end{center}
     \caption{VMD-based cross channel nucleon exchange model for $ N\gamma$ (left panel) and $ \gamma N$ (right panel) TDAs; dashed circles contain a typical LO graph for the nucleon electromagnetic form factor in perturbative QCD; the rectangle contains the cross channel nucleon contribution into $ N\gamma$ TDAs. The crossed circle depicts the $\gamma \to V_T $ vertex.}
\label{Fig_TDA_VMD}
\end{figure}

We assume the validity of the collinear factorized description for the-near backward
$\gamma^* N \to V_T N'$ and $\gamma^* N \to \gamma N'$
in terms of $VN$ ({\it resp.} $\gamma N$) TDAs and nucleon DAs.
This results in the following VMD-based model for $\gamma N$ TDAs (see Fig.~\ref{Fig_TDA_VMD}):
\be
&&
\{V, \, A\}_{\Upsilon}^{\gamma N}=\frac{e}{f_{\rho}} \{V, \, A\}_{\Upsilon}^{\rho N}+\frac{e}{f_{\omega}} \{V, \, A\}_{\Upsilon}^{\omega N}+\frac{e}{f_{\phi}} \{V, \, A\}_{\Upsilon}^{\phi N}, \ \ \text{with} \ \
\Upsilon= 1{\cal E}, \, 1T, \, 2{\cal E}, \, 2T;
\nn \\
&&
T_{\Upsilon}^{\gamma N}=\frac{e}{f_{\rho}} T_{\Upsilon}^{\rho N}+\frac{e}{f_{\omega}} T_{\Upsilon}^{\omega N}+\frac{e}{f_{\phi}} T_{\Upsilon}^{\phi_{T} N}, \ \ \text{with} \ \
\Upsilon= 1{\cal E}, \, 1T, \, 2{\cal E}, \, 2T, \, 3{\cal E}, \, 3T, 4{\cal E}; \, 4T;
\label{VMD_based_gN}
\ee

\subsection{Backward DVCS and TCS amplitudes in the cross-channel nucleon exchange approximation}
For our phenomenological applications
 we employ the cross-channel nucleon exchange model
for nucleon-to-vector meson TDAs
described in Sec.~5.1.3 of Ref.~\cite{Pire:2021hbl}.
This model has recently  seen a success in the description
of the near-backward $\omega$-meson electroproduction
\cite{JeffersonLabFp:2019gpp,Gayoso:2021rzj}.
The VMD-assumption
(\ref{VMD_based_gN})
combined with the cross channel nucleon exchange model for
$VN$ TDAs  results in great simplifications for the calculation
of convolution of resulting $\gamma N$ TDAs and nucleon DAs with hard scattering kernels.

The integral convolutions
$\mathcal{I}^{(k)}_{\rm\,  bDVCS}$, $k=1,\,3,\,4,\,5$,
within our model turn out to be  linear combinations of
$\mathcal{I}^{(k)}$s, $k=1,\,3,\,4,\,5$, corresponding to different vector mesons:
\be
&&
\mathcal{I}^{(1)}_{\rm\,  bDVCS}(\xi, \Delta^2) \Big|_{N(940)}=
\sum_{V=\rho, \omega, \phi} \frac{e}{f_{V}}
{\cal I}_0 \frac{K_{1 {\cal E}}^{VN}(\xi, \Delta^2, G^V_{VNN},G^T_{VNN})}{\xi}; \nn \\ &&
\mathcal{I}^{(3)}_{\rm\,  bDVCS}(\xi, \Delta^2) \Big|_{N(940)}=
\sum_{V=\rho, \omega, \phi} \frac{e}{f_{V}}
{\cal I}_0 \frac{K_{1 T}^{VN}(\xi, \Delta^2, G^V_{VNN},G^T_{VNN})+K_{2 {\cal E}}^{VN}(\xi, \Delta^2, G^V_{VNN},G^T_{VNN})}{\xi}; \nn \\ &&
\mathcal{I}^{(4)}_{\rm\,  bDVCS}(\xi, \Delta^2) \Big|_{N(940)}=
\sum_{V=\rho, \omega, \phi} \frac{e}{f_{V}} {\cal I}_0 \frac{K_{2 T}^{VN}(\xi, \Delta^2, G^V_{VNN},G^T_{VNN})}{\xi}; \nn \\ &&
\mathcal{I}^{(5)}_{\rm\,  bDVCS}(\xi, \Delta^2) \Big|_{N(940)}= -\sum_{V=\rho, \omega, \phi}  \frac{e}{f_{V}} {\cal I}_0 \frac{K_{2 {\cal E}}^{VN}(\xi, \Delta^2, G^V_{VNN},G^T_{VNN})}{\xi}\,. 
\label{Conv_I_bDVCS}
\ee
Here $G^V_{VNN}$ and $G^T_{VNN}$ stand for the vector and tensor couplings of vector mesons to nucleons
(see {\it e.g.} Ref.~\cite{Semenov-Tian-Shansky:2007yeh} for
the explicit form of the $VNN$ effective vertices).
The explicit expressions for the functions
$K_{1 {\cal E}}^{VN}$,
$K_{2 {\cal E}}^{VN}$,
$K_{1 T}^{VN}$,
$K_{2 T}^{VN}$,
read \cite{Pire:2021hbl}
\be
&&
K_{1 {\mathcal{E}}}^{VN}( \xi, \Delta^2)= \frac{f_N}{\Delta^2-m_N^2} \left( G^V_{VNN} \frac{2\xi(1-\xi)}{1+\xi} +
G^T_{VNN}\,  \xi
\left(
\frac{2 \xi }{1+\xi }-\frac{ \Delta^2}{m_N^2}
\right)
\right); \nn \\ &&
K_{2 {\mathcal{E}}}^{VN}( \xi, \Delta^2)=\frac{f_N}{\Delta^2-m_N^2} \left( G^V_{VNN}(-2 \xi) + G^T_{VNN}\xi \right);
\nn \\ &&
K_{1 T}( \xi, \Delta^2)= \frac{f_N}{\Delta^2-m_N^2} \left( -G^V_{VNN} \frac{2 \xi  (1+3 \xi)}{1-\xi}   \right);
\nn \\ &&
K_{2 T}( \xi, \Delta^2)= \frac{f_N}{\Delta^2-m_N^2} \left( G^T_{VNN}   \frac{\xi (1+\xi)}{1-\xi} \right).
\label{Def_K_factors}
\ee

The  constant ${\cal I}_0$  is the same constant occurring in the
leading order perturbative QCD description of proton electromagnetic form factor $F_1^p(Q^2)$
\cite{Chernyak:1987nv}:
\be
Q^4 F_1^p(Q^2)= \frac{(4 \pi \alpha_s)^2 f_N^2}{54}{\cal I}_0.
\label{pQCDF1}
\ee
The value  of ${\cal I}_0$
depends on the input phenomenological solutions for nucleon DAs\footnote{
For the Chernyak-Ogloblin-Zhitnitsky (COZ) input nucleon DA \cite{Chernyak:1987nv}
$|{\cal I}_0| \Big|_{\rm COZ}=144844$;
an input DA with a shape close to the asymptotic form
results
in a negligibly small value of  ${\cal I}_0$,
while a compromise Braun-Lenz-Wittmann  Next-to-Next-to-Leading-Order (BLW NNLO) solution \cite{Lenz:2009ar}  provides
$|{\cal I}_0| \Big|_{\rm BLW \, NNLO}=45470$.
}.
In the $\Delta_T \to 0$ limit
only the contributions of
$3$ TDAs
$V_{1 {{\cal E} }}^{\gamma N}$, $A_{1 {{\cal E} }}^{\gamma N}$, and
$T_{1 {{\cal E} }}^{\gamma N}+T_{2 {{\cal E} }}^{\gamma N}$
into the integral convolution
$\mathcal{I}^{(1)}_{\rm\,  bDVCS}$
turn out to be relevant%
\footnote{
The combination $T_{1 {{\cal E} }}^{N \gamma }-T_{2 {{\cal E} }}^{N \gamma }$ contributes only to a helicity-$\frac{3}{2}$ baryonic final state, and would thus be accessible in the reaction
$\gamma^* N \to \gamma \Delta(h=\pm 3/2)$.
}.

Our next step is to trace the effect of crossing from bDVCS to bTCS.
It turns out that the denominators $D_\alpha$ of hard kernels read as
\be
D_\alpha(x_i,\xi)=\frac{1}{2\xi} \tilde{D}_\alpha \left(\frac{x_i}{2 \xi} \right);
\ee
and TDAs within cross channel nucleon exchange model
are also functions of $\frac{x_i}{2 \xi}$ ratios
times some coefficients depending  on $\xi$.
{\it E.g.} for the $1 {\cal E}$ set:
\be
&&
\Big\{ V_{1 {\cal E}}^{VN}, \,  A_{1 {\cal E}}^{VN}, \,
T_{1 {\cal E}}^{VN},  T_{2 {\cal E}}^{VN} \Big\}
(x_1, x_2, x_3, \xi, \Delta^2) \Big|_{N(940)} \nn \\ && =
\Theta_{\rm ERBL}(x_1,x_2,x_3) \frac{1}{(2 \xi)^2}
K_{1 {\cal E}}^{VN}( \xi, \Delta^2)
\Big\{
V^p, \, A^p, \, -T^p,\,-T^p
\Big\} \left(  \frac{x_1}{2 \xi}, \frac{x_2}{2 \xi}, \frac{x_3}{2 \xi}  \right),
\ee
where $\Theta_{\rm ERBL}(x_1,x_2,x_3)\equiv  \prod_{k=1}^3 \theta(0 \le x_k \le 2 \xi)$ ensures that the integration with hard kernels
is restricted to the ERBL-like support region.
Therefore, up to an irrelevant phase factor, the effect of the crossing
(\ref{Crossing_prescription})
allowing us to express the integral convolutions
$\mathcal{J}^{(k)}_{\rm\,  bTCS}(\xi, \Delta^2) \Big|_{N(940)}$
of Eq.~(\ref{TransAmpl_squared_Cross})
through
$\mathcal{I}^{(k)}_{\rm\,  bDVCS}(\xi, \Delta^2) \Big|_{N(940)}$
of Eq.~(\ref{TransAmpl_squared})
within our VMD-based model
is reduced  to a mere change of the sign of $\xi$ in the arguments of
$K_{\Upsilon}^{VN}$ factors in Eq.~(\ref{Conv_I_bDVCS}):
\be
&&
\mathcal{J}^{(1)}_{\rm\,  bTCS}(\xi, \Delta^2) \Big|_{N(940)} \doteq
\sum_{V=\rho, \omega, \phi} \frac{e}{f_{V}}
{\cal I}_0 \frac{K_{1 {\cal E}}^{VN}(-\xi, \Delta^2, G^V_{VNN},G^T_{VNN})}{\xi}; \nn \\ &&
\mathcal{J}^{(3)}_{\rm\,  bTCS}(\xi, \Delta^2) \Big|_{N(940)} \doteq
\sum_{V=\rho, \omega, \phi} \frac{e}{f_{V}}
{\cal I}_0 \frac{K_{1 T}^{VN}(-\xi, \Delta^2, G^V_{VNN},G^T_{VNN})+K_{2 {\cal E}}^{VN}(-\xi, \Delta^2, G^V_{VNN},G^T_{VNN})}{\xi}; \nn \\ &&
\mathcal{J}^{(4)}_{\rm\,  bTCS}(\xi, \Delta^2) \Big|_{N(940)} \doteq
\sum_{V=\rho, \omega, \phi} \frac{e}{f_{V}} {\cal I}_0 \frac{K_{2 T}^{VN}(-\xi, \Delta^2, G^V_{VNN},G^T_{VNN})}{\xi}; \nn \\ &&
\mathcal{J}^{(5)}_{\rm\,  bTCS}(\xi, \Delta^2) \Big|_{N(940)}\doteq  -\sum_{V=\rho, \omega, \phi}  \frac{e}{f_{V}} {\cal I}_0 \frac{K_{2 {\cal E}}^{VN}(-\xi, \Delta^2, G^V_{VNN},G^T_{VNN})}{\xi}\,. 
\label{Conv_I_bTCS}
\ee

Note, that the change of
sign of the imaginary part of integral convolutions makes no effect since
in our  cross-channel nucleon exchange model with solely ERBL-like
support, the imaginary part of the amplitude is zero.

\begin{figure}[H]
 \begin{center}
 \includegraphics[width=0.47\textwidth]{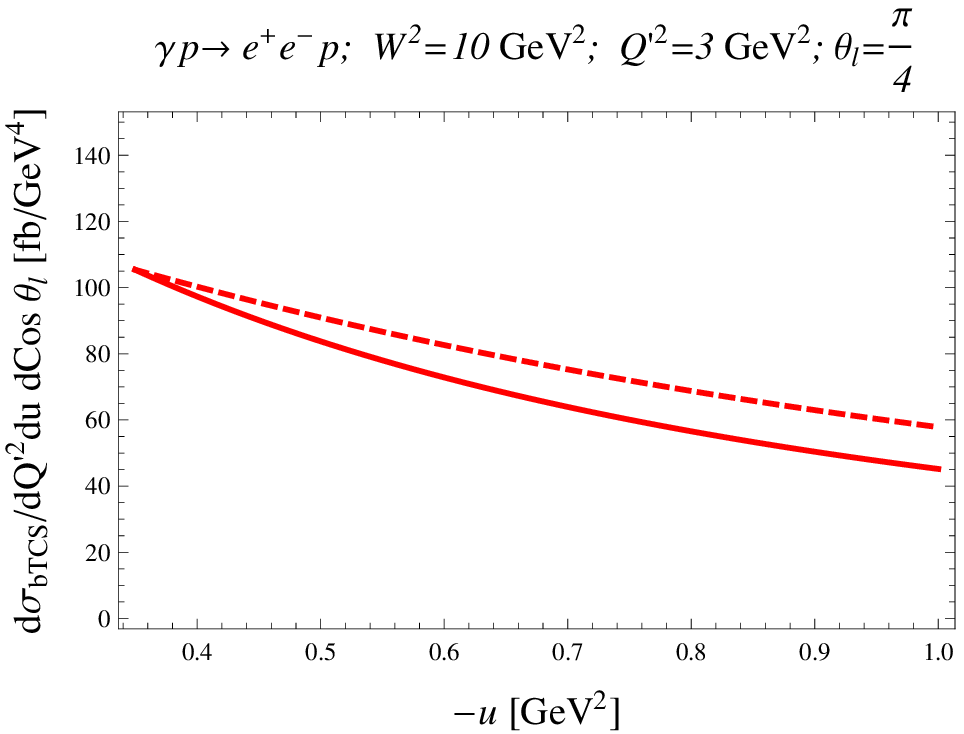}
  \includegraphics[width=0.47\textwidth]{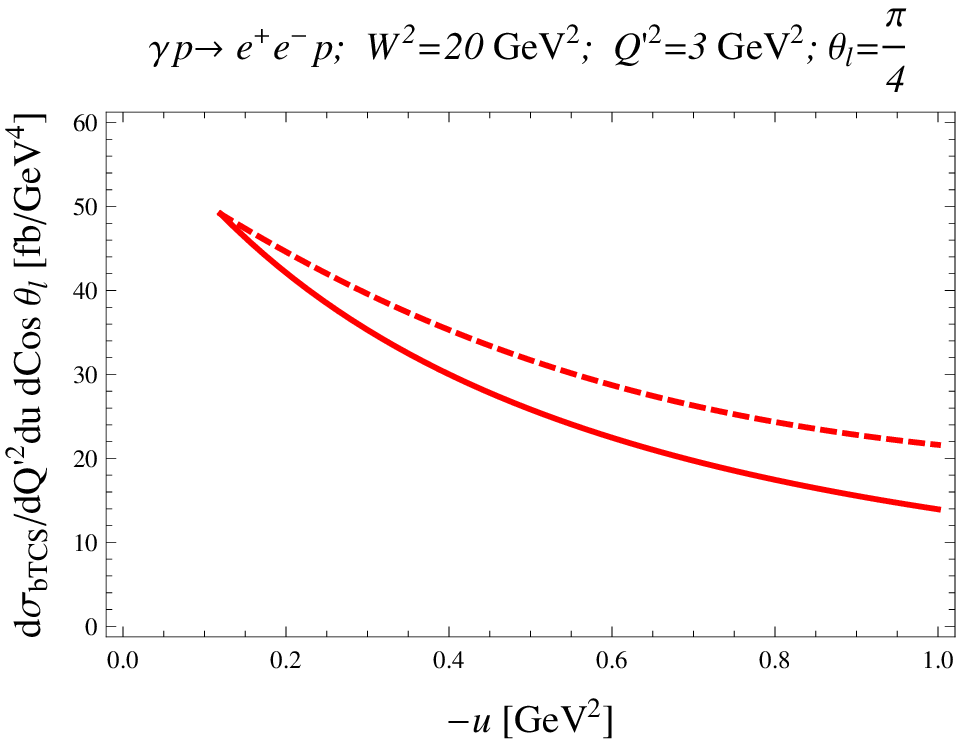}
   \includegraphics[width=0.47\textwidth]{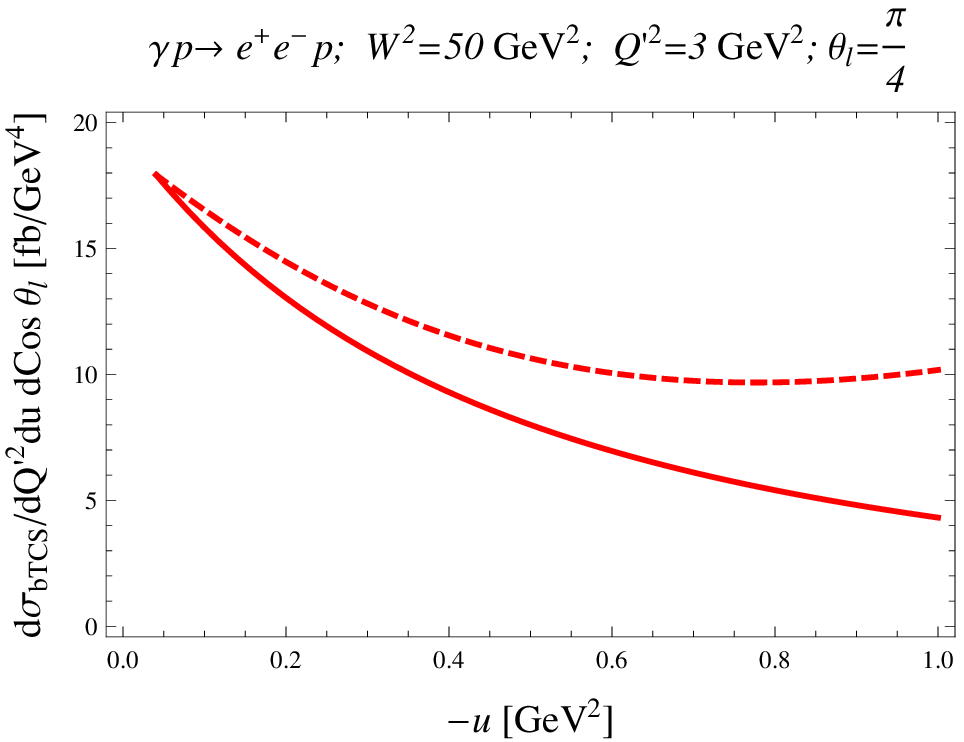}
    \includegraphics[width=0.47\textwidth]{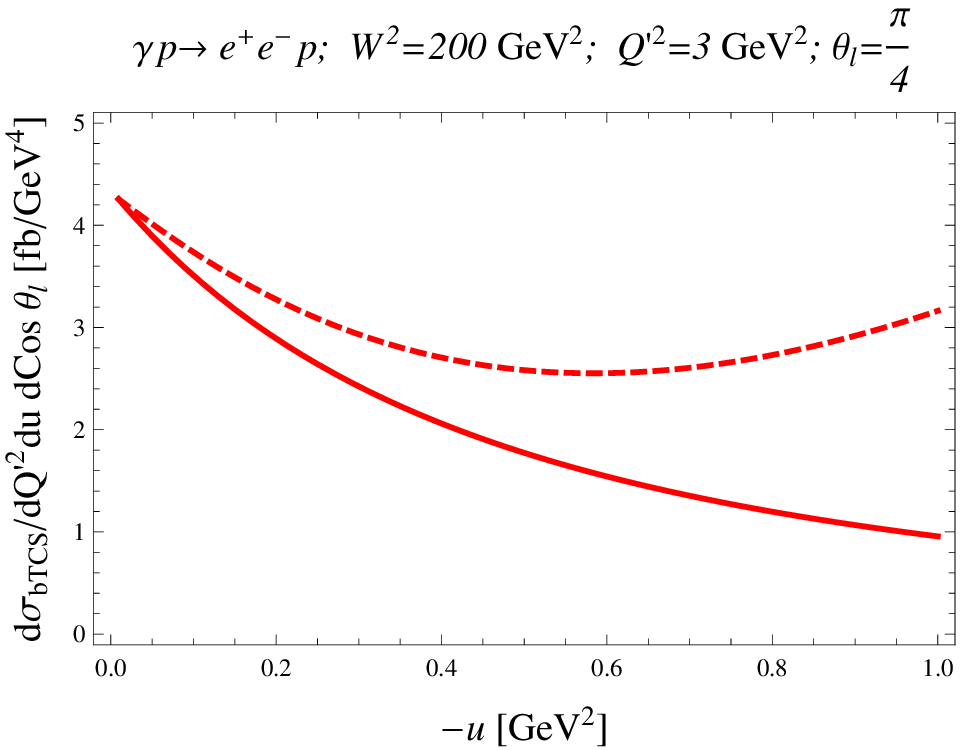}
     \includegraphics[width=0.47\textwidth]{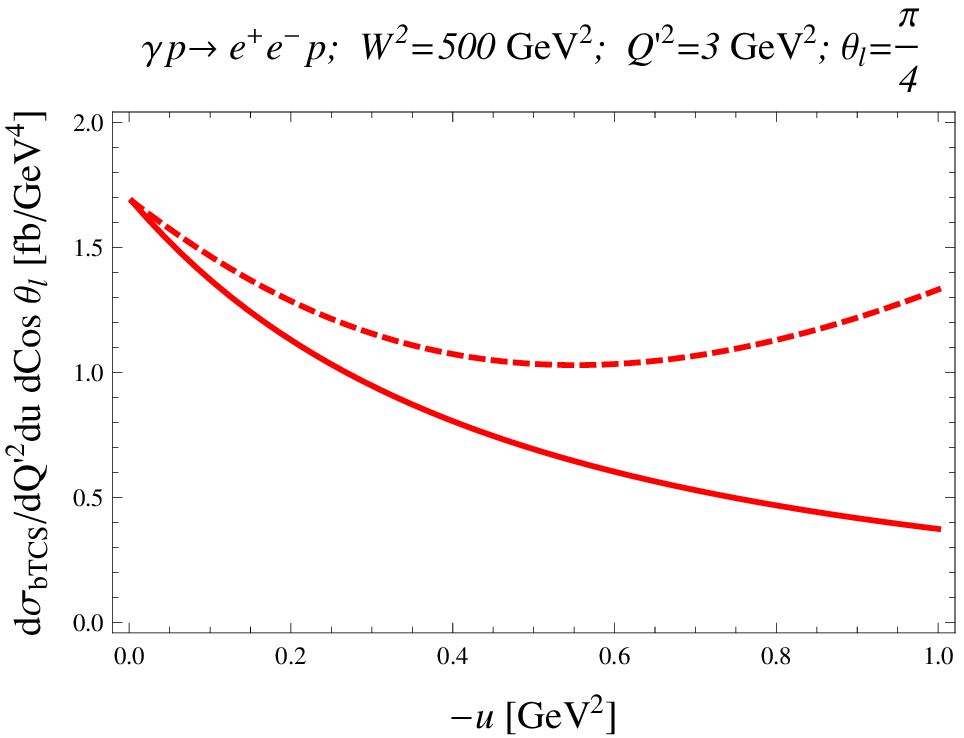}
 \end{center}
  \caption{ The near-backward  $\gamma p \to p e^+ e^-$ scattering cross section
(\ref{CS_main_formula}) for several values of $W^2$; $Q'^2=3$ GeV$^2$; and lepton polar angle $\theta_\ell=\frac{\pi}{4}$ as a function of $-u$ from the minimal value $-u_0$ up to $1$ GeV$^2$.
COZ solution is employed as input phenomenological
solution for nucleon DAs. Solid lines show the contribution of
$4$ $N \gamma$ TDAs
$V_{1 {\cal E}}^{N \gamma}, \,  A_{1 {\cal E}}^{N \gamma}, \,
T_{1,2 {\cal E}}^{N \gamma}$ relevant in the strictly backward $\Delta_T=0$ limit.
Dashed lines account also for the contributions of
$V_{2 {\cal E}, 2T}^{N \gamma}, \,  A_{2 {\cal E} 2T}^{N \gamma}, \,
T_{3 {\cal E}, 4 {\cal E}, 2T,3T}^{N \gamma}$ vanishing  for $\Delta_T=0$.
}
\label{Fig_CSVMD}
\end{figure}


\subsection{Backward TCS cross section estimate}

In Fig.~\ref{Fig_CSVMD} we show the estimates of the
near-backward  $\gamma p \to p e^+ e^-$ scattering cross section
(\ref{CS_main_formula})
within the reaction mechanism involving $N \gamma$ TDAs relying on the VMD assumption
employing COZ input nucleon DAs. We  take into account only the contribution of $\omega(782)$.
Note that, as clearly can be seen from eq.~(\ref{pQCDF1}), the cross section of
 $\gamma p \to p e^+ e^-$
within our model turns to be proportional to the square of
the perturbative QCD nucleon electromagnetic form factor squared since
\be
\overline{|{\cal M}_{N \gamma \to N' \ell^+ \ell^-}|^2}
\sim Q'^2 \left(F_1(Q'^2)\right)^2.
\ee
The kinematic factors occurring from
Eqs.~(\ref{Def_K_factors}) bring additional dependence on $Q'^2$ and $W^2$
through the skewness variable $\xi$ (\ref{Xi_collinear}).

Our model must be read as a very rough but plausible order of magnitude estimate%
\footnote{Experimental data on the timelike nucleon form factors (see for instance the review \cite{Tomasi-Gustafsson:2015zza}) indicate that at moderate $Q^2$ they are a factor around $2$ greater than the corresponding spacelike ones. We do not include this factor of $2$ in our present estimates.}
since the cross channel
nucleon exchange model for $\omega N$ TDAs saw some success in the description
of hard exclusive backward $\omega$-production data \cite{JeffersonLabFp:2019gpp}.
We employ \footnote{The issue of $\rho$- and  $\omega$-meson vector and tensor couplings remains rather controversial in the literature
(see {\it e.g.} Table~2 of Ref.~\cite{Meissner:1997qt} for a summary).}
the set of $G^{V,T}_{\omega NN}$ couplings Bonn~2000
\cite{Machleidt:2000ge}. The cross section is plotted for several values of  $W^2$; $Q'^2=3$ GeV$^2$; and lepton polar angle $\theta_\ell=\frac{\pi}{4}$ as a function of $-u$ from the minimal value $-u_0$ up to $1$ GeV$^2$.
Solid lines show the contribution of
$4$ $N \gamma$ TDAs
$V_{1 {\cal E}}^{N \gamma}, \,  A_{1 {\cal E}}^{N \gamma}, \,
T_{1{\cal E},2 {\cal E}}^{N \gamma}$ relevant in the strictly backward $\Delta_T=0$ limit.
Dashed lines account also for the contributions of
$V_{2 {\cal E}, 2T}^{N \gamma}, \,  A_{2 {\cal E}, 2T}^{N \gamma}, \,
T_{3 {\cal E}, 4 {\cal E}, 2T,3T}^{N \gamma}$ vanishing  for $\Delta_T=0$.
The first two plots of Fig.~\ref{Fig_CSVMD} correspond to JLab@12 GeV and JLab@24 GeV kinematical conditions while subsequent plots correspond to the future EIC and EicC kinematic range.

It worth emphasizing that the signature of the onset of the partonic picture of backward TCS is thus not to be searched for in the $Q'^2$ dependence of the amplitude, which is the same as in a nucleon exchange hadronic model, provided one explains the $Q^2$ behavior of the nucleon form factor in this model. The real signature of the TDA framework is the fact that this $Q'^2$ behavior appears at fixed value of $\xi$.

The polarization signature ( namely that the lepton pair comes from a transversely polarized virtual photon) is also an important signal of this onset. Although we present our estimates at a definite value of $\theta_l$, it is easy to integrate them in a definite $\theta_l$ bin since we know,
see Eq.~(\ref{M2_leading_tw}), the $1+\cos^2 \theta_l$ shape of this angular dependence.

Let us stress that the TDA picture is on no account
limited to a simple cross channel
nucleon exchange TDA model we employed here.
Further studies with a more realistic  model for
$\gamma N$ and $N \gamma$
TDAs can bring more than our present very crude cross section estimates.
In particular, advanced TDA models with complete domain of definition
can provide predictions for polarization observables sensitive to TDAs outside the ERBL-like region. For example, a sizable $Q'^2$-independent  transverse target single spin asymmetry (see Sec.~7.2.1 of \cite{Pire:2021hbl}) for bTCS can be seen as a strong
evidence in favor of validity of the factorization mechanism.
Moreover, the sign of this asymmetry is predicted to be different for bTCS and bDVCS,
which provides a useful additional cross check.

\section{Cross sections with an electron beam}
\label{Sec_CS_eN}

Quasi-real photoproduction generates a large fraction of the data recorded at an electron facility, such as JLab, or a future electron-ion collider \cite{AbdulKhalek:2021gbh,Anderle:2021wcy}. The cross-section is then calculated with the help of the Weizs\"acker-Williams distribution \cite{Kessler:1975hh,Frixione:1993yw} as:
\begin{equation}
\sigma_{eN} = \int_{x_{\min}}^1 dx \sigma_{\rm TCS}(x) f(x)\,,
    \label{WW}
\end{equation}
where  $x$ is the fraction of energy lost by the incoming electron in the target rest frame: $x= \frac{s_{\gamma N} - m_N^2}{s_{e N} - m_N^2}$ and $f(x)$, the photon distribution in the electron, reads
\begin{equation}
   f(x)=\frac{\alpha_{\rm em}}{2 \pi}
\left\{2 m_e^2 x
\left(\frac{1}{Q^2_{\rm max}} -\frac{1-x}{m_e^2 x^2}  \right)
+ \frac{\left((1-x)^2+1\right)
\ln \frac{Q^2_{\rm max}(1-x)}{m_e^2 x^2}}x
\right\}.
\end{equation}
Here $m_e$ is the electron mass and $Q^2_{\rm max}$ is the typical maximal value of the virtuality of the exchanged photon, which depends on the detection system ({\it e.g.}  $0.15$~GeV$^{2}$ as in the CLAS12 experiment at JLab \cite{CLAS:2021lky}, which we
use for our numerical estimates).

\begin{figure}[H]
 \begin{center}
 \includegraphics[width=0.7\textwidth]{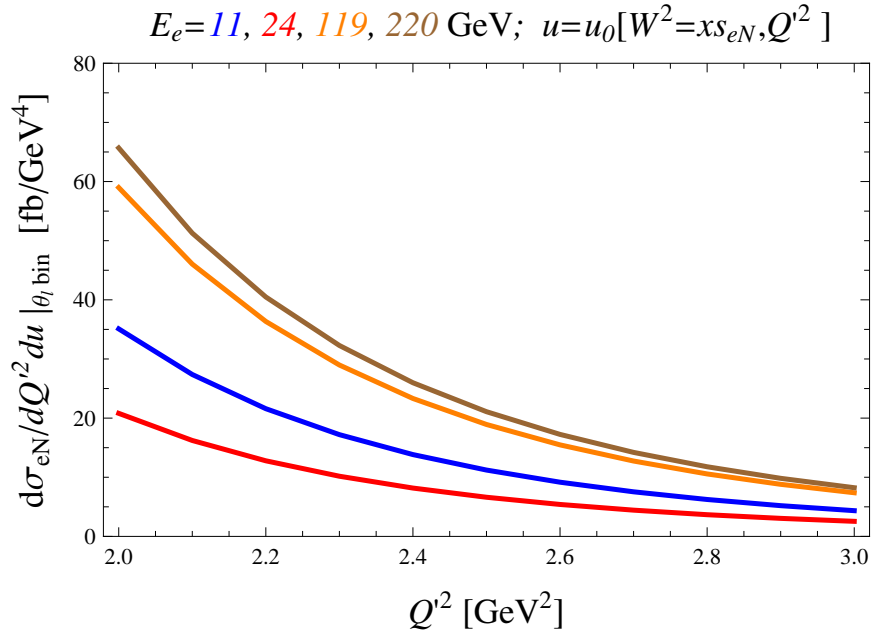}
 \end{center}
  \caption{ The electron-nucleon cross section (\ref{Def_CS_eN}) integrated over
  $\theta_\ell$
  as a function of ${Q'}^2$ for $u=u_0$ (exactly backward TCS)
    at electron beam energies $E_e$ corresponding (from down to up) to kinematical conditions of JLab@12, JLab@24 and of EIC and EicC.
     We employ the VMD-based model for $\gamma N$ TDAs and account for the contribution of $\omega(782)$ meson. We use
the set of
$G^{V,T}_{\omega NN}$
couplings Bonn~2000
\cite{Machleidt:2000ge} and COZ solution for nucleon DAs
as phenomenological input.
    }
\label{CS_EIcC}
\end{figure}

In Figure~\ref{CS_EIcC} we present the  $\frac{d\sigma_{eN}}{du dQ'^2 d \cos \theta_\ell}$ cross section integrated over the $\theta_\ell \in \left[\frac{\pi}{8}; \, \frac{7\pi}{8}\right]$ bin:
\be
&&
\frac{d\sigma_{eN}}{d {Q'}^2 du } \Big|_{\theta_\ell \, {\rm bin}}
\equiv
\int_{\theta_\ell \, {\rm bin}} d \theta_\ell \sin \theta_\ell
\frac{d\sigma_{eN}}{du dQ'^2 d \cos \theta_\ell}
= \int_{x_{\min}}^1 dx \int_{\theta_\ell \, {\rm bin}} d \theta_\ell \sin \theta_\ell
 \frac{d \sigma_{\rm TCS}(x)}{d {Q'}^2du d \cos \theta_\ell} f(x). \nn \\ &&
 \label{Def_CS_eN}
\ee
The cut in $\theta_\ell$ is intended to stay away from the narrow BH peak around
$\theta_\ell=0$
(see Sec.~\ref{Sec_BH}) and the symmetric peak around $\theta_\ell=\pi$. We define  $x_{\min}= \frac{s_{\gamma N \min} - m_N^2}{s_{e N} - m_N^2}$ and set
$s_{\gamma N  \min} =8$~GeV$^2$.

This plot clearly shows the feasibility of the experiment in an accelerator of high luminosity (say above $10$~fb$^{-1}$ year$^{-1}$) and a good lepton detector system for identifying the final lepton pair against a meson pair. Let us stress that our oversimplified model for $N\gamma$ TDAs is rather conservative:
one may well expect more source of electromagnetic radiation inside a nucleon when a hard probe has extracted its baryonic core than in a quite nucleon. Moreover the VMD picture of photon-nucleon interaction has not been tested in any comparable  experimental setting as the backward TCS one.

\section{Conclusions and Outlook}
\label{Sec_Concl}

In this paper, we analyzed within the collinear QCD factorization framework a never previously discussed promising reaction, namely (quasi-)real photoproduction of a lepton pair in the backward region. This reaction may be analyzed in terms of  photon to nucleon TDAs the characteristic features of which we have described. Although no complete model of these non-perturbative quantities exists at the moment, we have argued that a simple-minded estimate of the cross-sections indicate that the experimental rates should be sufficient for a dedicated study at JLab to be fruitful.

Although we discussed for simplicity the case of electron-positron pair production, it is clear that our results apply also to $\mu^+ \mu^-$ pair production (with minor corrections related to the final lepton mass effect), which may be easier to  experimentally identify if the detector is instrumented with a muon detection system in the

Ultra-peripheral collisions (UPCs) at high energy hadron colliders are also a fertile source of photoproduction reactions \cite{Bertulani:2005ru, Baltz:2007kq}. It has already been noted \cite{Pire:2008ea, Pire:2009ev} that the intense quasi-real photon radiation generated by moving protons (and the even more intense one generated by nuclei) may trigger a new domain of experimental exploration of GPDs in the small skewness region through the study of near forward lepton pair production. Likewise, the near-backward TCS process that we study here may provide us with a new way to access photon-to-nucleon TDAs.

Experiments on a nuclear target will enlarge the study proposed here through a search for color transparency effects \cite{Jain:2022xzo} which may help to decide where the onset of the dominance of short distance physics hence the validity of collinear QCD factorization takes place.

In conclusion, the study of the inner content of the nucleon already benefits much from the high luminosity experiments at medium energy electron facilities through the promising first results on the extraction of GPDs and transverse momentum dependent (TMDs) distributions. More is still to come and the TDA portal - both for nucleon-to-meson and nucleon-to-photon cases
\cite{CLAS:2017rgp, CLAS:2020yqf, JeffersonLabFp:2019gpp,Li:2020nsk,Gayoso:2021rzj}
- opens completely new opportunities to access new hadronic matrix elements of quite fundamental quark trilocal operators on the light cone.

\section{Acknowledgements}
We acknowledge useful discussions with Pierre Chatagnon, Stefan Diehl, Garth Huber, Bill Li, Silvia Nicolai, and Justin Stevens. We thank the anonymous referee for her/his very careful reading and  help in finding a calculation mistake in the first version of the paper.

This work was supported in part by the European Union's Horizon 2020 research and innovation programme under Grant Agreement No. 824093 and by the LABEX P2IO.
The work of K.S. and A.S. is supported by the Foundation for
the Advancement of Theoretical Physics and Mathematics ``BASIS''.
The work of L.S. is supported by the grant 2019/33/B/ST2/02588 of the National Science Center in Poland.

\setcounter{section}{0}
\setcounter{equation}{0}
\renewcommand{\thesection}{\Alph{section}}
\renewcommand{\theequation}{\thesection\arabic{equation}}

\section{Near-Backward TCS Amplitude}
\label{App_A}

In this Appendix we present the explicit results for the near-backward
TCS amplitude. The expressions for the coefficients
$T_{\alpha}^{(k)}$, $k=1,3,4,5$
occurring in Eq.~(\ref{Def_I_k_convolutions})
result from the calculation of the familiar $21$ diagrams and
take the form of products of singular
hard kernels $D_\alpha \equiv  D_\alpha^{(k)}(x_1,x_2,x_3,\xi; \, y_1, y_2,y_3)$, originating from the partonic propagators,
times the combinations
$N_\alpha^{(k)} \equiv N_\alpha^{(k)}(x_1,x_2,x_3,\xi, \Delta^2; \, y_1, y_2,y_3)$
of
$N \gamma$ TDAs and nucleon DAs
arising in the numerators:
\begin{equation}
T_{\alpha}^{(k)}\equiv  D_\alpha \times N_\alpha^{(k)}.
\label{Def_T_alpha}
\end{equation}
Note that no summation over the repeating index $\alpha$ is assumed in
(\ref{Def_T_alpha}).

Up to an irrelevant phase factor, the expressions for
$ D_\alpha$  and $N_\alpha^{(k)}$, $k=1,3,4,5$
can be obtained from those listed%
\footnote{In Table~3
of Ref.~\cite{Pire:2021hbl} a wrong sign occurs for $A_{1 {\cal E}, 2 {\cal E}, 1n, 2n, 1T, 2T}$, and
$T_{4 {\cal E}}$ entries.  }
in Table~3
of Ref.~\cite{Pire:2021hbl}
with obvious replacement of nucleon-to-vector-meson ($VN$) TDAs with
appropriate photon-to-nucleon ($N \gamma$) TDAs and changing the sign of
$i0$ regularization prescription in the denominators of corresponding hard scattering kernels.
This change
mirrors the fact that we consider $\gamma^*$ in the final state rather
than in the initial state, as it is for backward vector meson
electroproduction.
We list the corresponding results below in Table~I.

\newpage
\begin{longtable}{|c|p{5.1cm} |c|}
\caption{14 of the 21 diagrams contributing to the hard-scattering amplitude with
their associated coefficient $T_\alpha^{(k)} \equiv D_\alpha \times N_\alpha^{(k)}$ (no summation over $\alpha$ assumed).    The seven first ones with $u$-quark
lines inverted are not drawn. The crosses represent the virtual-photon vertex. $Q_u$ and $Q_d$ stand for the quark charges.} \\
\hline
$\alpha$ &  \qquad \quad Diagram  & Numerators \\
        &  \qquad \qquad  $D_\alpha$  &  \\
        \nopagebreak
\hline
1 & \raisebox{-0.0cm}
{\includegraphics[height=1.5cm,clip=true]{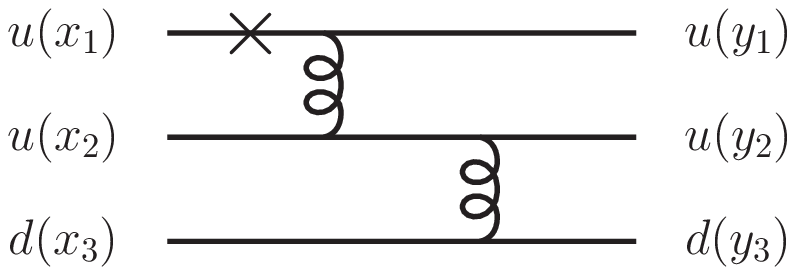}}
$\frac{Q_u (2\xi)^2}{(2\xi-x_{1}+i0)^2(x_{3}+i0)(1-y_{1})^2y_{3}}$ &
\begin{tabular}{p{0.85cm}|p{9.2cm}}
 $N_\alpha^{(1)}$ & $
-\left(V^p-A^p\right)
   \left(V_{1 {\cal E}}^{N \gamma}-A_{1 {\cal E}}^{N \gamma}\right)
   +
   2 T^{p} \left(T_{1 {\cal E}}^{N \gamma}+T_{2 {\cal E}}^{N \gamma}\right)
$
\\ \hline
\raisebox{-0.2cm}{  $N_\alpha^{(3)}$ } &
$
-\left(V^p-A^p\right)
   \left(V_{1 T}^{N \gamma}-A_{1 T}^{N \gamma}+V_{2 {\cal E}}^{N \gamma}-A_{2 {\cal E}}^{N \gamma}\right)
  $
\\
    &  $ +
   4 T^{p}  \left( T_{1 T}^{N \gamma}+T_{3 {\cal E}}^{N \gamma} + \frac{\Delta_T^2  }{2 M^2} T_{4T}^{N \gamma} \right)$  \\ \hline
  $N_\alpha^{(4)}$ &  $
-\left(V^p-A^p\right)
   \left(V_{2 T}^{N \gamma}-A_{2 T}^{N \gamma} \right)
   +
   2 T^{p}  \left( T_{2 T}^{N \gamma}+T_{3 T}^{N \gamma}   \right)
$ \\ \hline
 $N_\alpha^{(5)}$  &  $
\left(V^p-A^p\right)
   \left(V_{2 {\cal E}}^{N \gamma}-A_{2 {\cal E}}^{N \gamma} \right)
   -
   2 T^{p}  \left( T_{3 {\cal E}}^{N \gamma}+T_{4 {\cal E}}^{N \gamma}   \right)
$ \\
\end{tabular} \\
\hline
\hline
2 & \raisebox{-0.0cm}
{\includegraphics[height=1.5cm,clip=true]{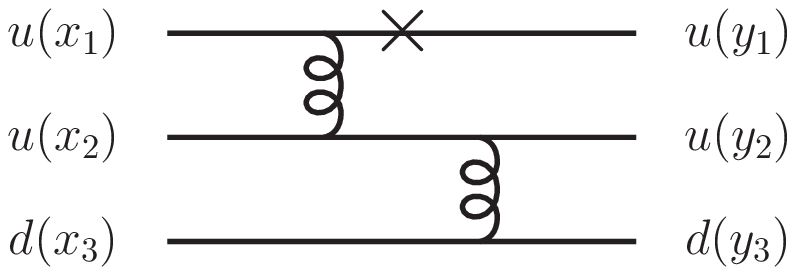}}
 &
\begin{tabular}{p{0.85cm}|p{9.2cm}}
 $N_\alpha^{(1)}$ &  \hspace{4.0cm}    $0$ \\ \hline
  $N_\alpha^{(3)}$ &  \hspace{4.0cm}    $0$ \\ \hline
 $N_\alpha^{(4)}$ &  \hspace{4.0cm}    $0$ \\ \hline
 $N_\alpha^{(5)}$ &  \hspace{4.0cm}    $0$ \\
\end{tabular} \\
\hline
\hline
3 & \raisebox{-0.0cm}
{\includegraphics[height=1.5cm,clip=true]{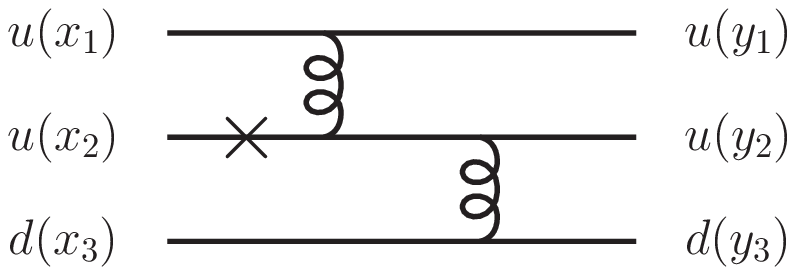}}
$\frac{Q_u (2 \xi)^2)}{(x_{1}+i0) (2\xi-x_{2}+i0)(x_{3}+i0)y_{1}(1-y_{1})y_{3}}$
 &
\begin{tabular}{p{0.85cm}|p{9.2cm}}
 $N_\alpha^{(1)}$ &  $-2 T^{p} \left(T_{1 {\cal E}}^{N \gamma}+T_{2 {\cal E}}^{N \gamma}\right)$ \\ \hline
  $N_\alpha^{(3)}$ & $-4 T^{p} \left(T_{1 T}^{N \gamma}+T_{3 {\cal E}}^{N \gamma}+\frac{\Delta_T^2}{2M^2}T_{4 T}^{N \gamma}\right)$ \\ \hline
 $N_\alpha^{(4)}$ & $-2 T^{p} \left(T_{2 T}^{N \gamma}+T_{3 T}^{N \gamma}\right)$ \\ \hline
 $N_\alpha^{(5)}$ & $2 T^{p} \left(T_{3 {\cal E}}^{N \gamma}+T_{4 {\cal E}}^{N \gamma}\right)$ \\ 
\end{tabular} \\
\hline
\hline
4 &  \raisebox{-0.0cm}
{\includegraphics[height=1.5cm,clip=true]{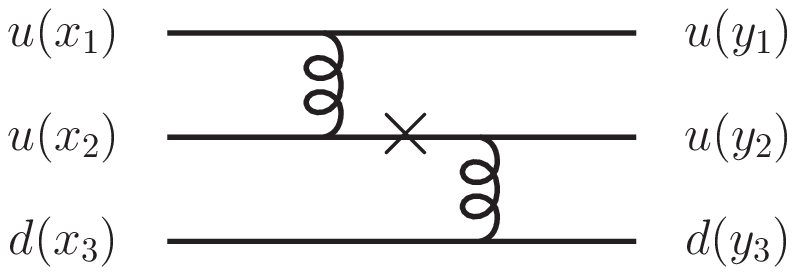}}
$\frac{Q_u (2\xi)^2}{(x_{1}+i0)
(2\xi-x_{3}+i0)(x_{3}+i0)y_{1}(1-y_{1})y_{3}}$
 &
\begin{tabular}{p{0.85cm}|p{9.2cm}}
 $N_\alpha^{(1)}$ &  $-\left(V^{p}-A^{p}\right) \left(V_{1 {\cal E} }^{N \gamma}-A_{1 {\cal E}}^{N \gamma}\right)$  \\ \hline
 $N_\alpha^{(3)}$ & $-\left(V^{p}-A^{p}\right) \left(V_{1 T }^{N \gamma}-A_{1 T}^{N \gamma}+ V_{2 {\cal E} }^{N \gamma}-A_{2 {\cal E}}^{N \gamma}\right)$ \\ \hline
 $N_\alpha^{(4)}$ & $-\left(V^{p}-A^{p}\right) \left(V_{2 T }^{N \gamma}-A_{2 T}^{N \gamma}\right)$ \\ \hline
 $N_\alpha^{(5)}$ & $\left(V^{p}-A^{p}\right) \left(V_{2 {\cal E} }^{N \gamma}-A_{2 {\cal E}}^{N \gamma}\right)$ \\
\end{tabular} \\
\hline
\hline
5 & \raisebox{-0.0cm}
{\includegraphics[height=1.5cm,clip=true]{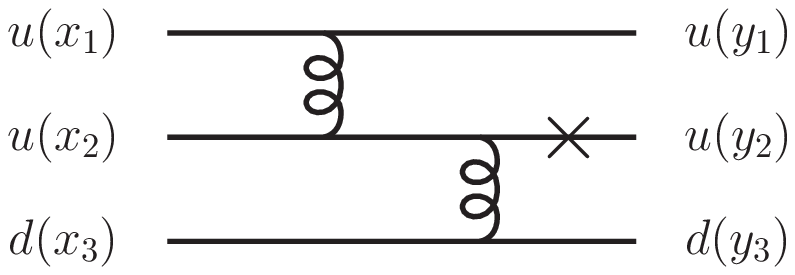}}
$\frac{Q_u (2\xi)^2}{(x_{1}+i0)(2\xi-x_{3}+i0)(x_{3}+i0)y_{1}(1-y_{2})y_{3}}$
 &
\begin{tabular}{p{0.85cm}|p{9.2cm}}
 $N_\alpha^{(1)}$ &  $\left(V^{p}+A^{p}\right) \left(V_{1 {\cal E} }^{N \gamma}+A_{1 {\cal E}}^{N \gamma}\right)$ \\  \hline
 $N_\alpha^{(3)}$ & $\left(V^{p}+A^{p}\right) \left(V_{1T }^{N \gamma}+A_{1 T}^{N \gamma}+V_{2 {\cal E} }^{N \gamma}+A_{2 {\cal E}}^{N \gamma}\right)$  \\ \hline
 $N_\alpha^{(4)}$ & $\left(V^{p}+A^{p}\right) \left(V_{2T }^{N \gamma}+A_{2 T}^{N \gamma} \right)$ \\ \hline
 $N_\alpha^{(5)}$ & $-\left(V^{p}+A^{p}\right) \left(V_{2 {\cal E} }^{N \gamma}+A_{2 {\cal E}}^{N \gamma}\right)$  \\ 
\end{tabular} \\
\hline
\hline
6 & \raisebox{-0.0cm}
{\includegraphics[height=1.5cm,clip=true]{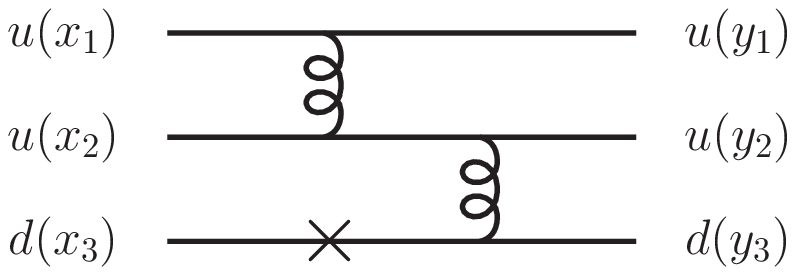}}
 &
\begin{tabular}{p{0.85cm}|p{9.2cm}}
 $N_\alpha^{(1)}$ &  \hspace{4.0cm} $0$ \\ \hline
 $N_\alpha^{(3)}$ & \hspace{4.0cm} $0$ \\ \hline
 $N_\alpha^{(4)}$ & \hspace{4.0cm} $0$ \\ \hline
 $N_\alpha^{(5)}$ & \hspace{4.0cm} $0$ \\
\end{tabular} \\
\hline
\hline
7 & \raisebox{-0.0cm}
{\includegraphics[height=1.5cm,clip=true]{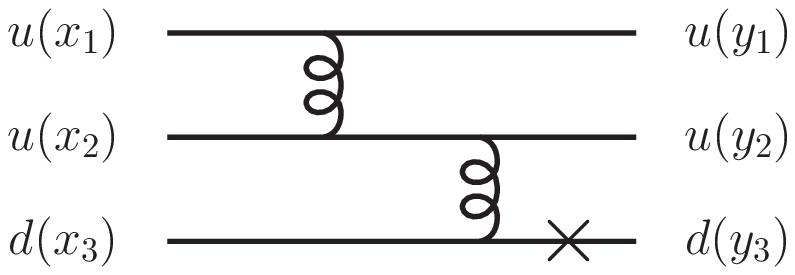}}
$\frac{Q_d(2\xi)^2}{(x_{1}+i0)(2\xi-x_{3}+i0)^2y_{1}(1-y_3)^2}$
 &
\begin{tabular}{p{0.85cm}|p{9.2cm}}
 $N_\alpha^{(1)}$ &    $-2 \left(V^{p} V_{1 {\cal E}}^{N \gamma}+A^{p}  A_{1 {\cal E}}^{N \gamma} \right)$ \\ \hline
  $N_\alpha^{(3)}$ &  $-2 \left(V^{p} (V_{1T}^{N \gamma}+V_{2 {\cal E}}^{N \gamma})+A^{p}  (A_{1T}^{N \gamma}+A_{2 {\cal E}}^{N \gamma}) \right)$ \\ \hline
 $N_\alpha^{(4)}$ & $-2 \left(V^{p} V_{2 T}^{N \gamma}+A^{p}  A_{2T}^{N \gamma} \right)$ \\ \hline
 $N_\alpha^{(5)}$ & $2 \left(V^{p} V_{2 {\cal E}}^{N \gamma}+A^{p}  A_{2 {\cal E}}^{N \gamma} \right)$  \\
\end{tabular} \\
\hline
\hline
8 & \raisebox{-0.0cm}
{\includegraphics[height=1.5cm,clip=true]{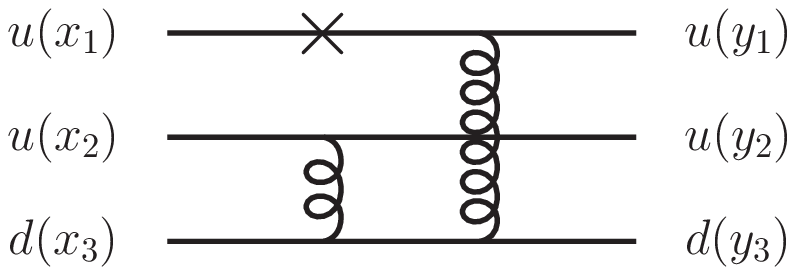}}
 &
\begin{tabular}{p{0.85cm}|p{9.2cm}}
 $N_\alpha^{(1)}$ &  \hspace{4.0cm} $0$ \\ \hline
  $N_\alpha^{(3)}$ & \hspace{4.0cm} $0$ \\ \hline
 $N_\alpha^{(4)}$ & \hspace{4.0cm} $0$ \\ \hline
 $N_\alpha^{(5)}$ & \hspace{4.0cm} $0$ \\
\end{tabular} \\
\hline
\hline
9 & \raisebox{-0.0cm}
{\includegraphics[height=1.5cm,clip=true]{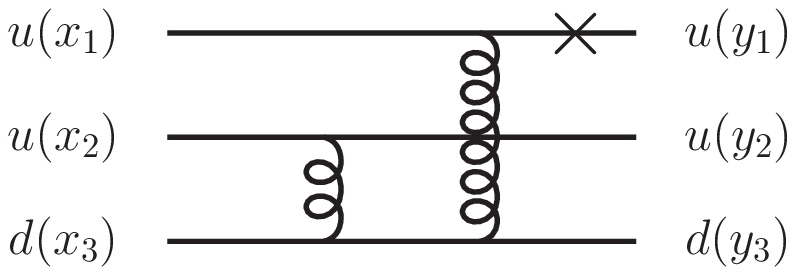}}
 \ \ \ \ \ \ \ \  \ \ \ \ \
$\frac{Q_u (2\xi)^2}{(2\xi-x_{1}+i0)^2(x_{2}+i0)(1-y_{1})^2y_{2}}$ &
\begin{tabular}{p{0.85cm}|p{9.2cm}}
 $N_\alpha^{(1)}$ & $
-\left(V^p-A^p\right)
   \left(V_{1 {\cal E}}^{N \gamma}-A_{1 {\cal E}}^{N \gamma}\right)
   +
   2 T^{p} \left(T_{1 {\cal E}}^{N \gamma}+T_{2 {\cal E}}^{N \gamma}\right)
$ \\ \hline
 \raisebox{-0.2cm}{  $N_\alpha^{(3)}$ } &
$
-\left(V^p-A^p\right)
   \left(V_{1 T}^{N \gamma}-A_{1 T}^{N \gamma}+V_{2 {\cal E}}^{N \gamma}-A_{2 {\cal E}}^{N \gamma}\right)
  $
\\
    &  $ +
   4 T^{p}  \left( T_{1 T}^{N \gamma}+T_{3 {\cal E}}^{N \gamma} + \frac{\Delta_T^2  }{2 M^2} T_{4T}^{N \gamma} \right)$  \\ \hline
  $N_\alpha^{(4)}$ &  $
-\left(V^p-A^p\right)
   \left(V_{2 T}^{N \gamma}-A_{2 T}^{N \gamma} \right)
   +
   2 T^{p}  \left( T_{2 T}^{N \gamma}+T_{3 T}^{N \gamma}   \right)
$ \\ \hline
 $N_\alpha^{(5)}$  &  $
\left(V^p-A^p\right)
   \left(V_{2 {\cal E}}^{N \gamma}-A_{2 {\cal E}}^{N \gamma} \right)
   -
   2 T^{p}  \left( T_{3 {\cal E}}^{N \gamma}+T_{4 {\cal E}}^{N \gamma}   \right)
$ \\
\end{tabular} \\
\hline
\hline
10 & \raisebox{-0.0cm}
{\includegraphics[height=1.5cm,clip=true]{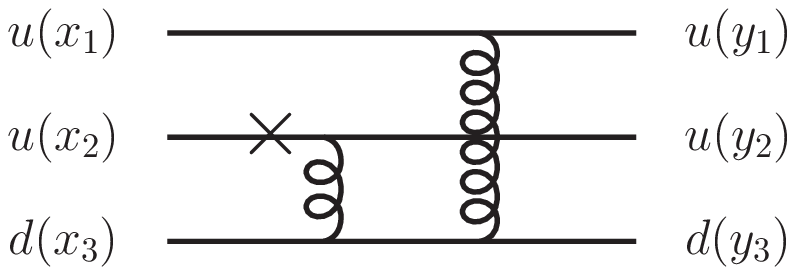}}
 \ \ \ \ \ \ \ \  \ \ \ \ \
$\frac{Q_u (2\xi)^2}{(x_{1}+i0)(2\xi-x_{2}+i0)^2y_{1}(1-y_{2})^2}$ &
\begin{tabular}{p{0.85cm}|p{9.2cm}}
 $N_\alpha^{(1)}$ & $
-\left(V^p-A^p\right)
   \left(V_{1 {\cal E}}^{N \gamma}-A_{1 {\cal E}}^{N \gamma}\right)
   +
   2 T^{p} \left(T_{1 {\cal E}}^{N \gamma}+T_{2 {\cal E}}^{N \gamma}\right)
$
\\ \hline
\raisebox{-0.2cm}{  $N_\alpha^{(3)}$ } &
$
-\left(V^p-A^p\right)
   \left(V_{1 T}^{N \gamma}-A_{1 T}^{N \gamma}+V_{2 {\cal E}}^{N \gamma}-A_{2 {\cal E}}^{N \gamma}\right)
  $
\\
    &  $ +
   4 T^{p}  \left( T_{1 T}^{N \gamma}+T_{3 {\cal E}}^{N \gamma} + \frac{\Delta_T^2  }{2 M^2} T_{4T}^{N \gamma} \right)$  \\ \hline
  $N_\alpha^{(4)}$ &  $
-\left(V^p-A^p\right)
   \left(V_{2 T}^{N \gamma}-A_{2 T}^{N \gamma} \right)
   +
   2 T^{p}  \left( T_{2 T}^{N \gamma}+T_{3 T}^{N \gamma}   \right)
$ \\ \hline
 $N_\alpha^{(5)}$  &  $
\left(V^p-A^p\right)
   \left(V_{2 {\cal E}}^{N \gamma}-A_{2 {\cal E}}^{N \gamma} \right)
   -
   2 T^{p}  \left( T_{3 {\cal E}}^{N \gamma}+T_{4 {\cal E}}^{N \gamma}   \right)
$  \\
\end{tabular} \\
\hline
\hline
11 & \raisebox{-0.0cm}
{\includegraphics[height=1.5cm,clip=true]{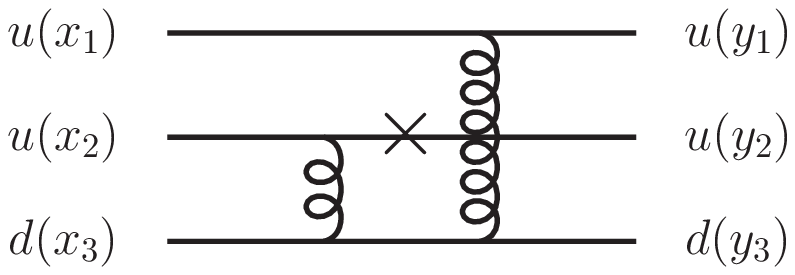}}
 &
\begin{tabular}{p{0.85cm}|p{9.2cm}}
 $N_\alpha^{(1)}$ &  \hspace{4.0cm} $0$ \\ \hline
 $N_\alpha^{(3)}$ & \hspace{4.0cm} $0$ \\ \hline
 $N_\alpha^{(4)}$ & \hspace{4.0cm} $0$ \\ \hline
 $N_\alpha^{(5)}$ & \hspace{4.0cm} $0$ \\
\end{tabular} \\
\hline
\hline
12 & \raisebox{-0.0cm}
{\includegraphics[height=1.5cm,clip=true]{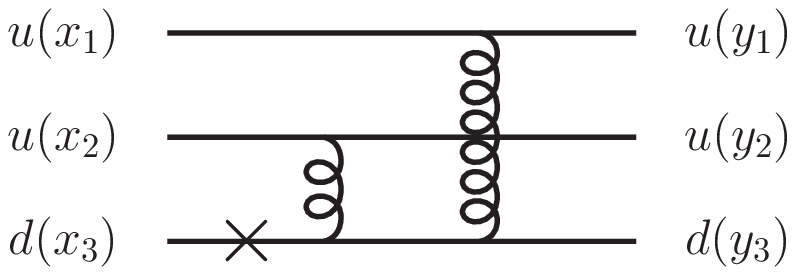}}
$\frac{Q_d (2 \xi)^2}{(x_{1}+i0)(x_{2}+i0)(2\xi-x_{3}+i0)y_{1}(1-y_{2})y_{2}}$
 &
\begin{tabular}{p{0.85cm}|p{9.2cm}}
 $N_\alpha^{(1)}$ &  $\left(V^{p}+A^{p}\right) \left( V_{1 {\cal E}}^{N \gamma}+A_{1 {\cal E}}^{N \gamma}\right)$
 \\ \hline
  $N_\alpha^{(3)}$ & $\left(V^{p}+A^{p}\right) \left( V_{1 {T}}^{N \gamma}+A_{1 {T}}^{N \gamma}+V_{2 {\cal E}}^{N \gamma}+A_{2 {\cal E}}^{N \gamma}\right)$  \\ \hline
 $N_\alpha^{(4)}$ & $\left(V^{p}+A^{p}\right) \left( V_{2 {T}}^{N \gamma}+A_{2 {T}}^{N \gamma}\right)$ \\ \hline
 $N_\alpha^{(5)}$ & $-\left(V^{p}+A^{p}\right) \left( V_{2 {\cal E}}^{N \gamma}+A_{2 {\cal E}}^{N \gamma}\right)$  \\
\end{tabular} \\
\hline
\hline
13 & \raisebox{-0.0cm}
{\includegraphics[height=1.5cm,clip=true]{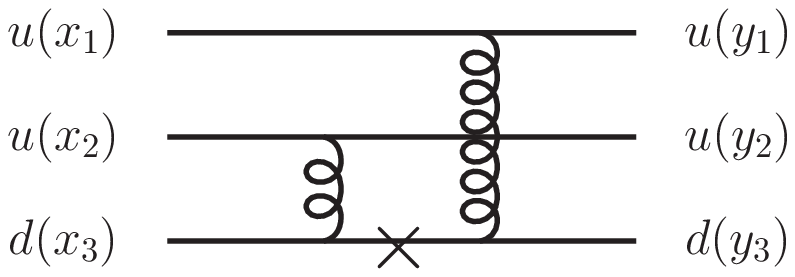}}
$\frac{Q_d (2 \xi)^2)}{(x_{1}+i0)(2\xi-x_{1}+i0)(x_{2}+i0)y_{1}(1-y_{2})y_{2}}$
 &
\begin{tabular}{p{0.85cm}|p{9.2cm}}
 $N_\alpha^{(1)}$ &  $2 T^{p} \left(T_{1 {\cal E}}^{N \gamma}+T_{2 {\cal E}}^{N \gamma}\right)$ \\ \hline
  $N_\alpha^{(3)}$ & $4 T^{p} \left(T_{1 T}^{N \gamma}+T_{3 {\cal E}}^{N \gamma}+\frac{\Delta_T^2}{2 M^2}T_{4 T}^{N \gamma}\right)$ \\ \hline
 $N_\alpha^{(4)}$ & $2 T^{p} \left(T_{2 T}^{N \gamma}+T_{3 T}^{N \gamma}\right)$ \\ \hline
 $N_\alpha^{(5)}$ & $-2 T^{p} \left(T_{3 {\cal E}}^{N \gamma}+T_{4 {\cal E}}^{N \gamma}\right)$
\end{tabular} \\
\hline
\hline
14 &  \raisebox{-0.0cm}
{\includegraphics[height=1.5cm,clip=true]{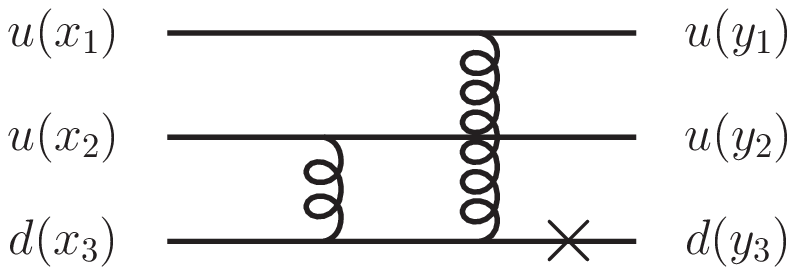}}
$\frac{Q_d (2\xi)^2}{(x_{1}+i0)(2\xi-x_{1}+i0)(x_{2}+i0)y_{1}y_{2}(1-y_{3})}$
 &
\begin{tabular}{p{0.85cm}|p{9.2cm}}
 $N_\alpha^{(1)}$ &  $\left(V^{p}-A^{p}\right) \left(V_{1 {\cal E} }^{N \gamma}-A_{1 {\cal E}}^{N \gamma}\right)$ \\ \hline
  $N_\alpha^{(3)}$ & $\left(V^{p}-A^{p}\right) \left(V_{1 T }^{N \gamma}-A_{1 T}^{N \gamma}+ V_{2 {\cal E} }^{N \gamma}-A_{2 {\cal E}}^{N \gamma}\right)$ \\ \hline
 $N_\alpha^{(4)}$ & $\left(V^{p}-A^{p}\right) \left(V_{2 T }^{N \gamma}-A_{2 T}^{N \gamma}\right)$ \\ \hline
 $N_\alpha^{(5)}$ & $-\left(V^{p}-A^{p}\right) \left(V_{2 {\cal E} }^{N \gamma}-A_{2 {\cal E}}^{N \gamma}\right)$
\end{tabular} \\
\hline
\end{longtable}

\setcounter{equation}{0}
\section{The set of Dirac structures for $\gamma N$ TDAs}
\label{App_B}

The set of  QED gauge invariant Dirac structures
$s_{\rho \tau, \chi}^{  \gamma N}$
occurring
in the definition of nucleon-to-photon TDAs are listed below in eqs.
(\ref{v_QED_gauge_inv})-(\ref{t_QED_gauge_inv}).
We adopt Dirac's ``hat'' notation
$\hat{l} \equiv l_\mu \gamma^\mu$;
$\sigma^{\mu\nu}= \frac{1}{2} [\gamma^\mu, \gamma^\nu]$; $\sigma_{p \mu} \equiv p^\lambda \sigma_{\lambda \mu}$;
$\sigma_{p {\Delta}_T} \equiv   p^\lambda \Delta_T^\mu \sigma_{\lambda \mu}$;
$C$
is the charge conjugation matrix; and
$U^+$
stands for the large component of the nucleon spinor
$
U^+(p_N,s_N)=  \hat{p} \hat{n}  U(p_N,s_N).
$

\be
&&
(v_{1 {{\cal E} }}^{\gamma N})_{\rho \tau, \, \chi}
= (\hat{p} C)_{\rho \tau}
\left[\left( \gamma_5 \hat{\cal E}^* U^{+}\right)_{\chi}-\frac{m_N}{1+\xi}
({\cal E}^* \cdot n)\left( \gamma_5 U^{+}\right)_{\chi}-\frac{2({\cal E}^* \cdot n)}{1-\xi}\left(\gamma_5 \hat{\Delta}_{T} U^{+}\right)_{\chi}\right]; \nn \\ &&
(v_{1 T}^{\gamma N})_{\rho \tau, \, \chi}
=\frac{1}{m_N}
\left[\left( {\cal E}^* \cdot \Delta_{T}\right)-\frac{2 \Delta_{T}^{2}}{1-\xi}
({\cal E}^* \cdot n)\right]
(\hat{p} C)_{\rho \tau}\left(\gamma_5 U^{+}\right)_{\chi}; \nn \\ &&
(v_{2 {\cal E}}^{\gamma N})_{\rho \tau, \, \chi}
=
\frac{1}{m_N}(\hat{p} C)_{\rho \tau}
\left[\left(\gamma_5 \sigma^{\Delta_{T} {\cal E}^*} U^{+}\right)_{\chi}-\frac{m_N
({\cal E}^* \cdot n)}{2(1+\xi)}\left(\gamma_5 \hat{\Delta}_{T} U^{+}\right)_{\chi}\right];
\nn \\ &&
(v_{2 T}^{\gamma N})_{\rho \tau, \, \chi}
=\frac{1}{m_N^2}
\left[\left( {\cal E}^* \cdot \Delta_{T}\right)-\frac{2 \Delta_{T}^{2}}{1-\xi}
({\cal E}^* \cdot n)\right]
(\hat{p} C)_{\rho \tau}\left(\gamma_5 \hat{\Delta}_T U^{+}\right)_{\chi}; \nn \\ &&
\label{v_QED_gauge_inv}
\ee
\be
&&
(a_{1 {{\cal E} }}^{\gamma N})_{\rho \tau, \, \chi}
= \left(\hat{p} \gamma^{5} C\right)_{\rho \tau}\left[\left(\hat{\cal E}^* U^{+}\right)_{\chi}-\frac{m_N}{1+\xi}({\cal E}^* \cdot n)\left( U^{+}\right)_{\chi}-\frac{2({\cal E}^* \cdot n)}{1-\xi}\left(  \hat{\Delta}_{T} U^{+}\right)_{\chi}\right]; \nn \\ &&
(a_{1 T}^{\gamma N})_{\rho \tau, \, \chi}=
\frac{1}{m_N}
\left[\left( {\cal E}^* \cdot \Delta_{T}\right)-\frac{2 \Delta_{T}^{2}}{1-\xi}
({\cal E}^* \cdot n)\right]
(\hat{p} \gamma^5 C)_{\rho \tau}\left(  U^{+}\right)_{\chi}; \nn \\ &&
(a_{2 {\cal E}}^{\gamma N})_{\rho \tau, \, \chi}
=
\frac{1}{m_N}(\hat{p}  \gamma^5 C)_{\rho \tau}
\left[\left( \sigma^{\Delta_{T} {\cal E}^*} U^{+}\right)_{\chi}-\frac{m_N
({\cal E}^* \cdot n)}{2(1+\xi)}\left(  \hat{\Delta}_{T} U^{+}\right)_{\chi}\right];
\nn \\ &&
(a_{2 T}^{\gamma N})_{\rho \tau, \, \chi}
=\frac{1}{m_N^2}
\left[\left( {\cal E}^* \cdot \Delta_{T}\right)-\frac{2 \Delta_{T}^{2}}{1-\xi}
({\cal E}^* \cdot n)\right]
(\hat{p}  \gamma^5 C)_{\rho \tau}\left(  \hat{\Delta}_T U^{+}\right)_{\chi}; 
\label{a_QED_gauge_inv}
\ee
\be
&&
(t_{1 {{\cal E} }}^{\gamma N})_{\rho \tau, \, \chi}
=\left(\sigma_{p \mu} C\right)_{\rho \tau}
\left[\left(\gamma_5 \sigma^{\mu {\cal E}^*} U^{+}\right)_{\chi}-\frac{m_N
({\cal E}^* \cdot n)}{2(1+\xi)}\left(\gamma^5 \gamma^{\mu} U^{+}\right)_{\chi}-\frac{2({\cal E}^* \cdot n)}{(1-\xi)}\left(\gamma^5 \sigma^{\mu \Delta_{T}} U^{+}\right)_{\chi}\right]; \nn \\ &&
(t_{1 T}^{\gamma N})_{\rho \tau, \, \chi}
=\frac{1}{m_N}\left[\left({\cal E}^* \cdot \Delta_{T}\right)-\frac{2 \Delta_{T}^{2}}{1-\xi}({\cal E}^* \cdot n)\right]\left(\sigma_{p \mu} C\right)_{\rho \tau}\left(\gamma^5 \gamma^{\mu} U^{+}\right)_{\chi}; \nn \\ &&
(t_{2 {{\cal E} }}^{\gamma N})_{\rho \tau, \, \chi}
=\left[\left(\sigma_{p {\cal E}^*} C\right)_{\rho \tau}-\frac{2({\cal E}^* \cdot n)}{(1-\xi)}\left(\sigma_{p \Delta_{T}} C\right)_{\rho \tau}\right]\left(\gamma^5 U^{+}\right)_{\chi}; \nn \\ &&
(t_{2 T}^{\gamma N})_{\rho \tau, \, \chi}
=\frac{1)}{m_N^{2}}
\left[\left({\cal E}^* \cdot \Delta_{T}\right)-\frac{2 \Delta_{T}^{2}}{1-\xi}
({\cal E}^* \cdot  n)\right]\left(\sigma_{p \mu} C\right)_{\rho \tau}\left(\gamma^5 \sigma^{\mu \Delta_{T}} U^{+}\right)_{\chi};
\nn \\ &&
(t_{3 {{\cal E} }}^{\gamma N})_{\rho \tau, \, \chi}
=
\frac{1}{m_N}\left(\sigma_{p \Delta_{T}} C\right)_{\rho \tau}\left[\left( \gamma^5 \hat{\cal E}^* U^{+}\right)_{\chi}-\frac{m_N ({\cal E}^* \cdot n)}{(1+\xi)}\left(\gamma^5 U^{+}\right)_{\chi}-\frac{2({\cal E}^* \cdot n)}{(1-\xi)}\left(\gamma^5 \hat{\Delta}_{T} U^{+}\right)_{\chi}\right]; \nn \\ &&
(t_{3 T}^{\gamma N})_{\rho \tau, \, \chi}
=
\frac{1}{m_N^{2}}\left[\left({\cal E}^* \cdot \Delta_{T}\right)-\frac{2 \Delta_{T}^{2}}{1-\xi}({\cal E}^* \cdot  n)\right]
\left(\sigma_{p \Delta_{T}} C\right)_{\rho \tau}\left(\gamma^5 U^{+}\right)_{\chi}; \nn \\ &&
(t_{4 {{\cal E} }}^{\gamma N})_{\rho \tau, \, \chi}
=\frac{1}{m_N}\left[\left(\sigma_{p {\cal E}^*} C\right)_{\rho \tau}-
\frac{2({\cal E}^* \cdot n)}{1-\xi}\left(\sigma_{p \Delta_{T}} C\right)_{\rho \tau}\right] \left(\gamma^5 \hat{\Delta}_{T} U^{+}\right)_{\chi}; \nn \\ &&
(t_{4 T}^{\gamma N})_{\rho \tau, \, \chi}
= \frac{1}{m_N^{3}}\left[\left({\cal E}^* \cdot \Delta_{T}\right)
-\frac{2 \Delta_{T}^{2}}{1-\xi}({\cal E}^* \cdot  n)\right]
\left(\sigma_{p \Delta_{T}} C\right)_{\rho \tau}\left(\gamma^5 \hat{\Delta}_{T} U^{+}\right)_{\chi}.
\label{t_QED_gauge_inv}
\ee

\bibliographystyle{elsarticle-num}
\bibliography{Backward_TCS}

\begin{thebibliography}{10}
\expandafter\ifx\csname url\endcsname\relax
  \def\url#1{\texttt{#1}}\fi
\expandafter\ifx\csname urlprefix\endcsname\relax\def\urlprefix{URL }\fi
\expandafter\ifx\csname href\endcsname\relax
  \def\href#1#2{#2} \def\path#1{#1}\fi

\bibitem{Muller:1994ses}
D.~M\"uller, D.~Robaschik, B.~Geyer, F.~M. Dittes, J.~Ho\v{r}ej\v{s}i, {Wave
  functions, evolution equations and evolution kernels from light ray operators
  of QCD}, Fortsch. Phys. 42 (1994) 101--141.
\newblock \href {http://arxiv.org/abs/hep-ph/9812448}
  {\path{arXiv:hep-ph/9812448}}, \href
  {http://dx.doi.org/10.1002/prop.2190420202}
  {\path{doi:10.1002/prop.2190420202}}.

\bibitem{Diehl:2003ny}
M.~Diehl, {Generalized parton distributions}, Phys. Rept. 388 (2003) 41--277.
\newblock \href {http://arxiv.org/abs/hep-ph/0307382}
  {\path{arXiv:hep-ph/0307382}}, \href
  {http://dx.doi.org/10.1016/j.physrep.2003.08.002}
  {\path{doi:10.1016/j.physrep.2003.08.002}}.

\bibitem{Belitsky:2005qn}
A.~V. Belitsky, A.~V. Radyushkin, {Unraveling hadron structure with generalized
  parton distributions}, Phys. Rept. 418 (2005) 1--387.
\newblock \href {http://arxiv.org/abs/hep-ph/0504030}
  {\path{arXiv:hep-ph/0504030}}, \href
  {http://dx.doi.org/10.1016/j.physrep.2005.06.002}
  {\path{doi:10.1016/j.physrep.2005.06.002}}.

\bibitem{Berger:2001xd}
E.~R. Berger, M.~Diehl, B.~Pire, {Time - like Compton scattering: Exclusive
  photoproduction of lepton pairs}, Eur. Phys. J. C 23 (2002) 675--689.
\newblock \href {http://arxiv.org/abs/hep-ph/0110062}
  {\path{arXiv:hep-ph/0110062}}, \href
  {http://dx.doi.org/10.1007/s100520200917} {\path{doi:10.1007/s100520200917}}.

\bibitem{Pire:2011st}
B.~Pire, L.~Szymanowski, J.~Wagner, {NLO corrections to timelike, spacelike and
  double deeply virtual Compton scattering}, Phys. Rev. D 83 (2011) 034009.
\newblock \href {http://arxiv.org/abs/1101.0555} {\path{arXiv:1101.0555}},
  \href {http://dx.doi.org/10.1103/PhysRevD.83.034009}
  {\path{doi:10.1103/PhysRevD.83.034009}}.

\bibitem{Moutarde:2013qs}
H.~Moutarde, B.~Pire, F.~Sabatie, L.~Szymanowski, J.~Wagner, {Timelike and
  spacelike deeply virtual Compton scattering at next-to-leading order}, Phys.
  Rev. D 87~(5) (2013) 054029.
\newblock \href {http://arxiv.org/abs/1301.3819} {\path{arXiv:1301.3819}},
  \href {http://dx.doi.org/10.1103/PhysRevD.87.054029}
  {\path{doi:10.1103/PhysRevD.87.054029}}.

\bibitem{Mueller:2012sma}
D.~M\"uller, B.~Pire, L.~Szymanowski, J.~Wagner, {On timelike and spacelike
  hard exclusive reactions}, Phys. Rev. D 86 (2012) 031502.
\newblock \href {http://arxiv.org/abs/1203.4392} {\path{arXiv:1203.4392}},
  \href {http://dx.doi.org/10.1103/PhysRevD.86.031502}
  {\path{doi:10.1103/PhysRevD.86.031502}}.

\bibitem{CLAS:2021lky}
P.~Chatagnon, et~al., {First Measurement of Timelike Compton Scattering}, Phys.
  Rev. Lett. 127~(26) (2021) 262501.
\newblock \href {http://arxiv.org/abs/2108.11746} {\path{arXiv:2108.11746}},
  \href {http://dx.doi.org/10.1103/PhysRevLett.127.262501}
  {\path{doi:10.1103/PhysRevLett.127.262501}}.

\bibitem{Pire:2021hbl}
B.~Pire, K.~Semenov-Tian-Shansky, L.~Szymanowski, {Transition distribution
  amplitudes and hard exclusive reactions with baryon number transfer}, Phys.
  Rept. 940 (2021) 2185.
\newblock \href {http://arxiv.org/abs/2103.01079} {\path{arXiv:2103.01079}},
  \href {http://dx.doi.org/10.1016/j.physrep.2021.09.002}
  {\path{doi:10.1016/j.physrep.2021.09.002}}.

\bibitem{Pire:2004ie}
B.~Pire, L.~Szymanowski, {Hadron annihilation into two photons and backward VCS
  in the scaling regime of QCD}, Phys. Rev. D 71 (2005) 111501.
\newblock \href {http://arxiv.org/abs/hep-ph/0411387}
  {\path{arXiv:hep-ph/0411387}}, \href
  {http://dx.doi.org/10.1103/PhysRevD.71.111501}
  {\path{doi:10.1103/PhysRevD.71.111501}}.

\bibitem{Lansberg:2006uh}
J.~P. Lansberg, B.~Pire, L.~Szymanowski, {Backward DVCS and Proton to Photon
  Transition Distribution Amplitudes}, Nucl. Phys. A 782 (2007) 16--23.
\newblock \href {http://arxiv.org/abs/hep-ph/0607130}
  {\path{arXiv:hep-ph/0607130}}, \href
  {http://dx.doi.org/10.1016/j.nuclphysa.2006.10.014}
  {\path{doi:10.1016/j.nuclphysa.2006.10.014}}.

\bibitem{Lansberg:2007se}
J.~P. Lansberg, B.~Pire, L.~Szymanowski, {Production of a pion in association
  with a high-$Q^2$ dilepton pair in antiproton-proton annihilation at
  GSI-FAIR}, Phys. Rev. D 76 (2007) 111502.
\newblock \href {http://arxiv.org/abs/0710.1267} {\path{arXiv:0710.1267}},
  \href {http://dx.doi.org/10.1103/PhysRevD.76.111502}
  {\path{doi:10.1103/PhysRevD.76.111502}}.

\bibitem{Lansberg:2012ha}
J.~P. Lansberg, B.~Pire, K.~Semenov-Tian-Shansky, L.~Szymanowski, {Accessing
  baryon to meson transition distribution amplitudes in meson production in
  association with a high invariant mass lepton pair at GSI-FAIR with
  $\overline {P}ANDA$}, Phys. Rev. D 86 (2012) 114033, [Erratum: Phys.Rev.D 87,
  059902 (2013)].
\newblock \href {http://arxiv.org/abs/1210.0126} {\path{arXiv:1210.0126}},
  \href {http://dx.doi.org/10.1103/PhysRevD.86.114033}
  {\path{doi:10.1103/PhysRevD.86.114033}}.

\bibitem{Pire:2016gut}
B.~Pire, K.~Semenov-Tian-Shansky, L.~Szymanowski, {Backward charmonium
  production in $\pi N$ collisions}, Phys. Rev. D 95~(3) (2017) 034021.
\newblock \href {http://arxiv.org/abs/1611.07234} {\path{arXiv:1611.07234}},
  \href {http://dx.doi.org/10.1103/PhysRevD.95.034021}
  {\path{doi:10.1103/PhysRevD.95.034021}}.

\bibitem{Pire:2019nwa}
B.~Pire, K.~Semenov-Tian-Shansky, L.~Szymanowski, {Nucleon-to-meson transition
  distribution amplitudes in impact parameter space}, PoS LC2019 (2019) 012.
\newblock \href {http://arxiv.org/abs/1912.05165} {\path{arXiv:1912.05165}},
  \href {http://dx.doi.org/10.22323/1.374.0012}
  {\path{doi:10.22323/1.374.0012}}.

\bibitem{Chernyak:1987nv}
V.~L. Chernyak, A.~A. Ogloblin, I.~R. Zhitnitsky, {Calculation of Exclusive
  Processes With Baryons}, Yad. Fiz. 48 (1988) 1398--1409.
\newblock \href {http://dx.doi.org/10.1007/BF01557664}
  {\path{doi:10.1007/BF01557664}}.

\bibitem{Chernyak:1984bm}
V.~L. Chernyak, I.~R. Zhitnitsky, {Nucleon Wave Function and Nucleon
  Form-Factors in QCD}, Nucl. Phys. B 246 (1984) 52--74.
\newblock \href {http://dx.doi.org/10.1016/0550-3213(84)90114-7}
  {\path{doi:10.1016/0550-3213(84)90114-7}}.

\bibitem{Borodulin:2017pwh}
V.~I. Borodulin, R.~N. Rogalyov, S.~R. Slabospitskii, {CORE 3.1 (COmpendium of
  RElations, Version 3.1)}\href {http://arxiv.org/abs/1702.08246}
  {\path{arXiv:1702.08246}}.

\bibitem{Pire:2015kxa}
B.~Pire, K.~Semenov-Tian-Shansky, L.~Szymanowski, {QCD description of backward
  vector meson hard electroproduction}, Phys. Rev. D 91~(9) (2015) 094006.
\newblock \href {http://arxiv.org/abs/1503.02012} {\path{arXiv:1503.02012}},
  \href {http://dx.doi.org/10.1103/PhysRevD.91.094006}
  {\path{doi:10.1103/PhysRevD.91.094006}}.

\bibitem{Hakioglu:1991pn}
T.~Hakioglu, M.~D. Scadron, {Vector meson dominance, one loop order quark
  graphs, and the chiral limit}, Phys. Rev. D 43 (1991) 2439--2442.
\newblock \href {http://dx.doi.org/10.1103/PhysRevD.43.2439}
  {\path{doi:10.1103/PhysRevD.43.2439}}.

\bibitem{Schildknecht:2005xr}
D.~Schildknecht, {Vector meson dominance}, Acta Phys. Polon. B 37 (2006)
  595--608.
\newblock \href {http://arxiv.org/abs/hep-ph/0511090}
  {\path{arXiv:hep-ph/0511090}}.

\bibitem{Dumbrajs:1983jd}
O.~Dumbrajs, R.~Koch, H.~Pilkuhn, G.~c. Oades, H.~Behrens, J.~j. De~Swart,
  P.~Kroll, {Compilation of Coupling Constants and Low-Energy Parameters. 1982
  Edition}, Nucl. Phys. B 216 (1983) 277--335.
\newblock \href {http://dx.doi.org/10.1016/0550-3213(83)90288-2}
  {\path{doi:10.1016/0550-3213(83)90288-2}}.

\bibitem{JeffersonLabFp:2019gpp}
W.~B. Li, et~al., {Unique Access to $u$-Channel Physics: Exclusive
  Backward-Angle Omega Meson Electroproduction}, Phys. Rev. Lett. 123~(18)
  (2019) 182501.
\newblock \href {http://arxiv.org/abs/1910.00464} {\path{arXiv:1910.00464}},
  \href {http://dx.doi.org/10.1103/PhysRevLett.123.182501}
  {\path{doi:10.1103/PhysRevLett.123.182501}}.

\bibitem{Gayoso:2021rzj}
C.~A. Gayoso, et~al., {Progress and opportunities in backward angle
  ($u$-channel) physics}, Eur. Phys. J. A 57~(12) (2021) 342.
\newblock \href {http://arxiv.org/abs/2107.06748} {\path{arXiv:2107.06748}},
  \href {http://dx.doi.org/10.1140/epja/s10050-021-00625-2}
  {\path{doi:10.1140/epja/s10050-021-00625-2}}.

\bibitem{Semenov-Tian-Shansky:2007yeh}
K.~M. Semenov-Tian-Shansky, A.~V. Vereshagin, V.~V. Vereshagin, {Bootstrap and
  the physical values of pi N resonance parameters}, Phys. Rev. D 77 (2008)
  025028.
\newblock \href {http://arxiv.org/abs/0706.3672} {\path{arXiv:0706.3672}},
  \href {http://dx.doi.org/10.1103/PhysRevD.77.025028}
  {\path{doi:10.1103/PhysRevD.77.025028}}.

\bibitem{Lenz:2009ar}
A.~Lenz, M.~Gockeler, T.~Kaltenbrunner, N.~Warkentin, {The Nucleon Distribution
  Amplitudes and their application to nucleon form factors and the $N \to
  \Delta$ transition at intermediate values of $Q^2$}, Phys. Rev. D 79 (2009)
  093007.
\newblock \href {http://arxiv.org/abs/0903.1723} {\path{arXiv:0903.1723}},
  \href {http://dx.doi.org/10.1103/PhysRevD.79.093007}
  {\path{doi:10.1103/PhysRevD.79.093007}}.

\bibitem{Tomasi-Gustafsson:2015zza}
E.~Tomasi-Gustafsson, {Proton electromagnetic form factors: present status and
  future perspectives at PANDA}, EPJ Web Conf. 95 (2015) 01015.
\newblock \href {http://dx.doi.org/10.1051/epjconf/20149501015}
  {\path{doi:10.1051/epjconf/20149501015}}.

\bibitem{Meissner:1997qt}
U.-G. Meissner, V.~Mull, J.~Speth, J.~W. van Orden, {Strange vector currents
  and the OZI rule}, Phys. Lett. B 408 (1997) 381--386.
\newblock \href {http://arxiv.org/abs/hep-ph/9701296}
  {\path{arXiv:hep-ph/9701296}}, \href
  {http://dx.doi.org/10.1016/S0370-2693(97)00828-9}
  {\path{doi:10.1016/S0370-2693(97)00828-9}}.

\bibitem{Machleidt:2000ge}
R.~Machleidt, {The High precision, charge dependent Bonn nucleon-nucleon
  potential (CD-Bonn)}, Phys. Rev. C 63 (2001) 024001.
\newblock \href {http://arxiv.org/abs/nucl-th/0006014}
  {\path{arXiv:nucl-th/0006014}}, \href
  {http://dx.doi.org/10.1103/PhysRevC.63.024001}
  {\path{doi:10.1103/PhysRevC.63.024001}}.

\bibitem{AbdulKhalek:2021gbh}
R.~Abdul~Khalek, et~al., {Science Requirements and Detector Concepts for the
  Electron-Ion Collider: EIC Yellow Report}\href
  {http://arxiv.org/abs/2103.05419} {\path{arXiv:2103.05419}}.

\bibitem{Anderle:2021wcy}
D.~P. Anderle, et~al., {Electron-ion collider in China}, Front. Phys. (Beijing)
  16~(6) (2021) 64701.
\newblock \href {http://arxiv.org/abs/2102.09222} {\path{arXiv:2102.09222}},
  \href {http://dx.doi.org/10.1007/s11467-021-1062-0}
  {\path{doi:10.1007/s11467-021-1062-0}}.

\bibitem{Kessler:1975hh}
P.~Kessler, {The Weizsacker-Williams Method and Similar Approximation Methods
  in Quantum Electrodynamics}, Acta Phys. Austriaca 41 (1975) 141--188.

\bibitem{Frixione:1993yw}
S.~Frixione, M.~L. Mangano, P.~Nason, G.~Ridolfi, {Improving the
  Weizsacker-Williams approximation in electron - proton collisions}, Phys.
  Lett. B 319 (1993) 339--345.
\newblock \href {http://arxiv.org/abs/hep-ph/9310350}
  {\path{arXiv:hep-ph/9310350}}, \href
  {http://dx.doi.org/10.1016/0370-2693(93)90823-Z}
  {\path{doi:10.1016/0370-2693(93)90823-Z}}.

\bibitem{Bertulani:2005ru}
C.~A. Bertulani, S.~R. Klein, J.~Nystrand, {Physics of ultra-peripheral nuclear
  collisions}, Ann. Rev. Nucl. Part. Sci. 55 (2005) 271--310.
\newblock \href {http://arxiv.org/abs/nucl-ex/0502005}
  {\path{arXiv:nucl-ex/0502005}}, \href
  {http://dx.doi.org/10.1146/annurev.nucl.55.090704.151526}
  {\path{doi:10.1146/annurev.nucl.55.090704.151526}}.

\bibitem{Baltz:2007kq}
A.~J. Baltz, {The Physics of Ultraperipheral Collisions at the LHC}, Phys.
  Rept. 458 (2008) 1--171.
\newblock \href {http://arxiv.org/abs/0706.3356} {\path{arXiv:0706.3356}},
  \href {http://dx.doi.org/10.1016/j.physrep.2007.12.001}
  {\path{doi:10.1016/j.physrep.2007.12.001}}.

\bibitem{Pire:2008ea}
B.~Pire, L.~Szymanowski, J.~Wagner, {Can one measure timelike Compton
  scattering at LHC?}, Phys. Rev. D 79 (2009) 014010.
\newblock \href {http://arxiv.org/abs/0811.0321} {\path{arXiv:0811.0321}},
  \href {http://dx.doi.org/10.1103/PhysRevD.79.014010}
  {\path{doi:10.1103/PhysRevD.79.014010}}.

\bibitem{Pire:2009ev}
B.~Pire, L.~Szymanowski, J.~Wagner, {Timelike Compton Scattering at LHC}, Acta
  Phys. Polon. Supp. 2 (2009) 373.
\newblock \href {http://arxiv.org/abs/0905.2056} {\path{arXiv:0905.2056}}.

\bibitem{Jain:2022xzo}
P.~Jain, B.~Pire, J.~P. Ralston, {The Status and Future of Color Transparency
  and Nuclear Filtering}, in: {The Future of Color Transparency and
  Hadronization Studies at Jefferson Lab and Beyond}, 2022.
\newblock \href {http://arxiv.org/abs/2203.02579} {\path{arXiv:2203.02579}}.

\bibitem{CLAS:2017rgp}
K.~Park, et~al., {Hard exclusive pion electroproduction at backward angles with
  CLAS}, Phys. Lett. B 780 (2018) 340--345.
\newblock \href {http://arxiv.org/abs/1711.08486} {\path{arXiv:1711.08486}},
  \href {http://dx.doi.org/10.1016/j.physletb.2018.03.026}
  {\path{doi:10.1016/j.physletb.2018.03.026}}.

\bibitem{CLAS:2020yqf}
S.~Diehl, et~al., {Extraction of Beam-Spin Asymmetries from the Hard Exclusive
  $\pi^+$ Channel off Protons in a Wide Range of Kinematics}, Phys. Rev. Lett.
  125~(18) (2020) 182001.
\newblock \href {http://arxiv.org/abs/2007.15677} {\path{arXiv:2007.15677}},
  \href {http://dx.doi.org/10.1103/PhysRevLett.125.182001}
  {\path{doi:10.1103/PhysRevLett.125.182001}}.

\bibitem{Li:2020nsk}
W.~B. Li, et~al., {Backward-angle Exclusive $\pi^0$ Production above the
  Resonance Region}\href {http://arxiv.org/abs/2008.10768}
  {\path{arXiv:2008.10768}}.

\end{thebibliography}
\end{document}